\documentclass{article}
\usepackage[margin=3cm]{geometry}

\usepackage{amsmath,amssymb,amsfonts}
\usepackage{mathrsfs} 
\usepackage{mathtools}

\usepackage{graphicx,color}
\usepackage[nocompress]{cite}

\newcommand{\ourG}{\mathbf{G}}
\newcommand{\ourB}{\mathbf{B}}
\newcommand{\ourC}{\mathbf{C}}

\DeclareMathOperator{\sech}{sech}
\DeclareMathOperator{\csch}{csch}
\DeclareMathOperator{\arctanh}{arctanh}

\numberwithin{equation}{section}

\usepackage{hyperref}
\hypersetup{colorlinks=true,allcolors=blue,urlcolor=cyan}

\title{Modified gravity: a unified approach to metric-affine models}
\author{Christian G. B\"ohmer\footnote{Email: c.boehmer@ucl.ac.uk} \ and
  Erik Jensko\footnote{Email: erik.jensko.19@ucl.ac.uk} \\
  Department of Mathematics, University College London, \\
  Gower Street, London WC1E 6BT, United Kingdom}

\date{26 January 2023} 

\pagecolor{white}

\begin{document}

\maketitle

\begin{abstract}
  The starting point of this work is the original Einstein action, sometimes called the Gamma squared action. Continuing from our previous results, we study various modified theories of gravity following the Palatini approach. The metric and the connection will be treated as independent variables leading to generalised theories which may contain torsion or non-metricity or both. Due to our particular approach involving the Einstein action, our setup allows us to formulate a substantial number of new theories not previously studied. Our results can be linked back to well-known models like Einstein-Cartan theory and metric-affine theories and also links to many recently studied modified gravity models. In particular we propose an Einstein-Cartan type modified theory of gravity which contains propagating torsion provided our function depends non-linearly on a boundary term. We also can state precise conditions for the existence of propagating torsion. Our work concludes with a brief discussion of cosmology and the role of cosmological torsion in our model. We find solutions with early-time inflation and late-time matter dominated behaviour. No matter sources are required to drive inflation and it becomes a purely geometrical effect. 
\end{abstract}

\clearpage

\tableofcontents

\clearpage

\section{Introduction}

The complete formulation of the Einstein field equations was followed, almost immediately, by a variational formulation due to Hilbert. Perhaps surprisingly, the simplest curvature scalar that can be defined on a Lorentzian manifold, namely the Ricci scalar, acts as the Lagrangian when deriving the Einstein field equations using the calculus of variations. The gravitational action based on this Lagrangian is the Einstein-Hilbert action. In many ways the Einstein-Hilbert action is somewhat unusual when compared to other field theories. The action contains second derivatives of the dynamical variables which are the components of the metric tensor. The resulting field equations are nonetheless of second order because it is possible to write all these second derivative terms in the form of a total derivative which will not contribute to the equations of motion.

When this total derivative or boundary term is subtracted from the Ricci scalar one finds an action quadratic in the Christoffel symbol components, which is known as the Einstein action~\cite{Pauli2000,Padmanabhan:2004xk,TP:2010}. While more natural than the Einstein-Hilbert action, when compared to other field theories, this action is no longer a coordinate scalar. It differs from a coordinate scalar by the boundary term which was subtracted. In our previous work~\cite{Boehmer:2021aji} we presented a comprehensive study of modified theories of gravity that are based on this decomposition of the Ricci scalar into two distinct terms. We often refer to those as the bulk and the boundary terms.

It is natural to extend this work in the context of metric-affine theories of gravity, see~\cite{Blagojevic:2013xpa,Hehl:1994ue}. By this we mean to treat the metric and the connection as independent variables, this is also often called the Palatini approach. When considering the Einstein-Hilbert action and assuming a manifold with curvature and torsion, one will naturally arrive at Einstein-Cartan theory~\cite{FH1976}, also see~\cite{Obukhov:2022khx} for a recent review. In this model matter is the source of curvature while spin acts as the source of torsion. A well-known feature of Einstein-Cartan theory is that torsion does not propagate, its field equation being algebraic. This means regions of spacetime without torsion sources cannot contain torsion.

The approach put forward in this work will allow us to formulate a new Einstein-Cartan type modified theory of gravity which contains propagating torsion. In fact, we show that Einstein-Cartan theory is the unique, minimally coupled, diffeomorphism invariant theory in which torsion does not propagate (for non-minimal coupling considerations see~\cite{Karananas:2021zkl,Rigouzzo:2022yan}). In passing we state the precise conditions for the existence of propagating torsion within our models, shown to be consistent with other theories with propagating torsion~\cite{Hojman:1978yz,Hojman:1979mg,Neville:1979rb,Sezgin:1979zf,Shapiro:2001rz,Baekler:2010fr,Baekler:2011jt}.

Our approach leads to a large variety of different modified theories of gravity, many of which have not been studied before. This is due to the decomposition of the Ricci scalar, which is the starting point of our considerations. This is different to many other modified theories of gravity, see for example~\cite{Lovelock:1971yv,Hehl:1994ue,Jacobson:2000xp,Sotiriou:2008rp,DeFelice:2010aj,Clifton:2011jh,Capozziello:2011et,RA2013,Koyama:2015vza, Cai:2015emx,Bahamonde:2015zma,Bahamonde:2016kba,Bahamonde:2017wwk,Krssak:2018ywd,Harko:2018ayt,BeltranJimenez:2019esp,Majid:2020,Bahamonde:2021gfp,CANTATA:2021ktz} for various publications covering a plethora of theories and approaches. Probably the most obvious difference between many previously studied models and ours is the decomposition of the Ricci scalar into diffeomorphism breaking terms. These terms appearing in this decomposition are not true coordinate scalars; they are not invariant under coordinate transformations in general, and we often refer to such terms as pseudo-scalars. We can therefore construct models that are distinctly different and cannot be directly linked back to previous studies due to the breaking of diffeomorphism invariance. Using, for example, a Born-Infeld type approach~\cite{Ferraro:2008ey,Fiorini:2009ux,Boehmer:2019uxv,Boehmer:2020hkn}, one could construct models that break diffeomorphism invariance on very small scales, where classical physics breaks down.

As an application of our construction we consider a modified Einstein-Cartan type theory in the context of flat Friedmann-Lemaitre-Robertson-Walker (FLRW) cosmology. Our model contains a simple quadratic non-linear term which is sufficient to give rise to propagating torsion. More importantly, torsion can exist without the need to introduce sources, similar to General Relativity where vacuum solution contains curvature. Torsion decays as the Universe expands, as one would expect. Most interestingly, we find early-time solutions where the Universe undergoes a period of accelerated expansion, i.e., inflation. Our solution does not require the introduction of sources to drive inflation, it becomes a purely geometrical effect. We carefully study the early-time and late-time asymptotic behaviour of our solution and compare our results with those of standard cosmology.

\subsection{Notation and definitions}

Following our previous work~\cite{Boehmer:2021aji}, Greek indices refer to spacetime objects and we use the signature $(-,+,+,+)$. An overline/overbar generally denotes the `full' object in the context of metric-affine spaces. This means $\overline{\Gamma}$ stands for the complete connection which contains the usual Levi-Civita part plus contributions from torsion and non-metricity. Where possible and practical we will follow the notation of Schouten~\cite{JS1954}. The covariant derivative $\nabla$ will always stand for the derivative with respect to the metric-affine connection. In the few instances where a different derivative is used, this is made explicit.

Torsion and non-metricity are defined in the usual ways
\begin{align}
  T^{\lambda}{}_{\mu \nu} &= 2 \overline{\Gamma}^{\lambda}_{[\mu \nu]} =
  \overline{\Gamma}^{\lambda}_{\mu \nu} - \overline{\Gamma}^{\lambda}_{\nu \mu} \,, \\
  Q_{\lambda \mu \nu} &= -\nabla_{\lambda} g_{\mu \nu}\,, \qquad
  Q_{\lambda}{}^{\mu \nu} = \nabla_{\lambda} g^{\mu \nu} \,.
\end{align}
We use the standard notation $A_{[\mu\nu]}=(A_{\mu\nu}-A_{\nu\mu})/2$ and $A_{(\mu\nu)}=(A_{\mu\nu}+A_{\nu\mu})/2$. The affine connection can be decomposed into its Levi-Civita part $\Gamma$ and torsion and non-metricity parts
\begin{align}
  \label{contorsion}
  \overline{\Gamma}{}^{\lambda}_{\mu \nu} =
  \Gamma{}^{\lambda}_{\mu \nu} + K_{\mu \nu}{}^{\lambda} \,, \qquad
  K_{\mu \nu}{}^{\lambda} = \frac{1}{2}  g^{\lambda \rho}
  \bigl( -T_{\{\nu \mu \rho \}} + Q_{\{ \mu \rho \nu \} } \bigr) \,,
\end{align}
where $K_{\mu \nu}{}^{\lambda}$ is known as the contortion tensor, and the Schouten bracket permutes indices according to
\begin{align}
  S_{\{\mu \nu \lambda \} } = S_{\mu \nu \lambda} - S_{\nu \lambda \mu } + S_{\lambda \mu \nu} \,.
\end{align}
The affine Riemann tensor is defined as
\begin{align}
  \overline{R}_{\mu \nu \lambda}{}^{\rho} = 2 \partial_{[\mu} \overline{\Gamma}{}^{\rho}_{\nu] \lambda} +
  2 \overline{\Gamma}{}^{\rho}_{[\mu | \sigma} \overline{\Gamma}{}^{\sigma}_{\nu] \lambda} \,,
\end{align}
from which we define the contractions $\overline{R}_{\nu \lambda} = \overline{R}_{\mu \nu \lambda}{}^{\mu}$ and $\overline{R} = g^{\mu \nu} \overline{R}_{\mu \nu}$ as the Ricci tensor and Ricci scalar respectively. Using the decomposition~(\ref{contorsion}) the total curvature can be written in terms of the Levi-Civita curvature and contortion
\begin{align}
  \overline{R}_{\mu \nu \lambda}{}^{\rho} = R_{\mu \nu \lambda}{}^{\rho} +
  2 K_{[\mu| \sigma}{}^{\rho} K_{\nu] \lambda}{}^{\sigma} +
  2 \stackrel{\{ \}}{\nabla}_{[\mu} K_{\nu] \lambda}{}^{\rho} \,,
\end{align}
with the covariant derivative here being with respect to the Levi-Civita connection.

\section{Modified metric-affine gravity}

This entire section contains the basic setup which underlies our work. At its heart is the decomposition of the metric-affine Ricci scalar and the subsequent split of its Levi-Civita part into a bulk term and a boundary term. As will be emphasised throughout this work, boundary terms play a crucial role in our approach and it is the particular way in which they are treated which gives rise to new modified theories of gravity.

\subsection{Affine curvature decomposition}

The Ricci scalar density for a general metric-affine connection $\overline{\Gamma}$ is given by
\begin{align}
  \sqrt{-g}\, \overline{R} = \sqrt{-g}g^{\mu\lambda}
  \bigl(\partial_\kappa \overline{\Gamma}_{\mu\lambda}^\kappa -
  \partial_\mu \overline{\Gamma}_{\kappa\lambda}^\kappa \bigr) + \sqrt{-g}g^{\mu\lambda}
  \bigl(\overline{\Gamma}_{\kappa\rho}^\kappa \overline{\Gamma}_{\mu\lambda}^\rho -
  \overline{\Gamma}_{\mu\rho}^\kappa \overline{\Gamma}_{\kappa\lambda}^\rho
  \bigr) \,.
\end{align}
In order to identify a suitable boundary term, we apply one integration by parts to each of the terms containing a first derivative of the connection. Written out explicitly, we have the two relations
\begin{align}
  \sqrt{-g}g^{\mu\lambda} \partial_\kappa \overline{\Gamma}_{\mu\lambda}^\kappa &=
  \partial_\kappa \bigl(\sqrt{-g}g^{\mu\lambda} \overline{\Gamma}_{\mu\lambda}^\kappa\bigr) -
  \sqrt{-g}\, \overline{\Gamma}_{\mu\lambda}^\kappa \partial_\kappa g^{\mu\lambda}
  +\frac{1}{2} \sqrt{-g} g_{\alpha\beta} g^{\mu\lambda} \overline{\Gamma}_{\mu\lambda}^\kappa
  \partial_\kappa g^{\alpha\beta} \,,\\
  \sqrt{-g}g^{\mu\lambda} \partial_\mu \overline{\Gamma}_{\kappa\lambda}^\kappa &=
  \partial_\mu \bigl(\sqrt{-g}g^{\mu\lambda} \overline{\Gamma}_{\kappa\lambda}^\kappa\bigr) -
  \sqrt{-g}\, \overline{\Gamma}_{\kappa\lambda}^\kappa \partial_\mu g^{\mu\lambda}
  +\frac{1}{2} \sqrt{-g} g_{\alpha\beta} g^{\mu\lambda} \overline{\Gamma}_{\kappa\lambda}^\kappa
  \partial_\mu g^{\alpha\beta} \,,
\end{align}
where the first term in each is a boundary term. Next, we can separate off these boundary terms and arrive at
\begin{multline}
  \sqrt{-g}\, \overline{R} =
  \partial_\kappa \bigl(\sqrt{-g}g^{\mu\lambda} \overline{\Gamma}_{\mu\lambda}^\kappa\bigr) -
  \partial_\mu \bigl(\sqrt{-g}g^{\mu\lambda} \overline{\Gamma}_{\kappa\lambda}^\kappa\bigr) \\ +
  \sqrt{-g}g^{\mu\lambda}
  \bigl(\overline{\Gamma}_{\kappa\rho}^\kappa \overline{\Gamma}_{\mu\lambda}^\rho -
  \overline{\Gamma}_{\mu\rho}^\kappa \overline{\Gamma}_{\kappa\lambda}^\rho\bigr) -
  \sqrt{-g}\, \overline{\Gamma}_{\mu\lambda}^\kappa \partial_\kappa g^{\mu\lambda} +
  \sqrt{-g}\, \overline{\Gamma}_{\kappa\lambda}^\kappa \partial_\mu g^{\mu\lambda} \\+
  \frac{1}{2} \sqrt{-g} g_{\alpha\beta} g^{\mu\lambda} \overline{\Gamma}_{\mu\lambda}^\kappa
  \partial_\kappa g^{\alpha\beta} -
  \frac{1}{2} \sqrt{-g} g_{\alpha\beta} g^{\mu\lambda} \overline{\Gamma}_{\kappa\lambda}^\kappa
  \partial_\mu g^{\alpha\beta} \,.
  \label{eqn:ac1}
\end{multline}

Introducing the notation $\overline{\ourB}$ to denote that boundary term and $\overline{\ourG}$ to denote the remaining bulk terms, we have the following decomposition
\begin{align}
  \sqrt{-g}\, \overline{R} = \sqrt{-g} (\overline{\mathbf{G}} + \mathbf{\overline{B}}) \,,
\end{align}
with $\overline{\mathbf{G}}$ and $\mathbf{\overline{B}}$ given by
\begin{align}
  \label{G}
  \overline{\mathbf{G}} &:= g^{\mu \lambda} \big( \overline{\Gamma}^{\kappa}_{\kappa \rho} \overline{\Gamma}^{\rho}_{\mu \lambda} - \overline{\Gamma}^{\kappa}_{\mu \rho} \overline{\Gamma}^{\rho}_{\kappa \lambda} \big)
  + \big(\overline{\Gamma}^{\mu}_{\mu \lambda} \delta^{\nu}_{\kappa} - \overline{\Gamma}^{\nu}_{\kappa \lambda} \big) \big( \partial_{\nu} g^{\kappa \lambda} - \frac{1}{2} g_{\alpha \beta} g^{\kappa \lambda} \partial_{\nu} g^{\alpha \beta} \big)\,, \\
  \label{B}
  \mathbf{\overline{B}} &:= \frac{1}{\sqrt{-g}} \partial_{\kappa} \big(\sqrt{-g}(g^{\mu \lambda}\overline{\Gamma}^{\kappa}_{\mu \lambda} - g^{\kappa \lambda} \overline{\Gamma}^{\mu}_{\mu \lambda})\big)\,.
\end{align}
The partial derivative terms in $\overline{\ourG}$ can be written in a slightly more succinct way 
\begin{align} \label{G2}
\overline{\ourG} = g^{\mu \lambda} \big( \overline{\Gamma}^{\kappa}_{\kappa \rho} \overline{\Gamma}^{\rho}_{\mu \lambda} - \overline{\Gamma}^{\kappa}_{\mu \rho} \overline{\Gamma}^{\rho}_{\kappa \lambda} \big)  - \frac{1}{2} \partial_{\lambda}g^{\mu \nu} \overline{E}_{\mu \nu}{}^{\lambda} \, ,
\end{align}
where the object $\overline{E}{}_{\mu \nu}{}^{\lambda}$ is defined by
\begin{align}
  \label{E}
  \overline{E}_{\mu \nu}{}^{\lambda} := 2 \overline{\Gamma}^{\lambda}_{(\mu \nu)} -
  2 \delta^{\lambda}_{(\mu} \overline{\Gamma}^{\rho}_{| \rho |\nu)} - g_{\mu \nu} g^{\kappa \rho} \overline{\Gamma}^{\lambda}_{\rho \kappa} +
  g_{\mu \nu} g^{\kappa \lambda} \overline{\Gamma}^{\rho}_{\rho \kappa} \,.
\end{align}
This will turn out to be useful when computing the field equations, and we will later return to the meaning of this non-tensorial, rank-three object. Let us also note that Eq.~(\ref{G2}) only determines the symmetric part of $\overline{E}_{\mu \nu}{}^{\lambda}$. Our first field equation is obtained by varying with respect to the metric and hence will only contain this very same symmetric object. However, when considering a generalised Einstein-Cartan type theory we will be able to also choose its skew-symmetric part in a particular way, see below.

Note that for the unique torsion-free, metric-compatible, Levi-Civita connection $\Gamma$, the bulk terms simplify to
\begin{align}
  g^{\mu \lambda} \bigl(\Gamma^{\kappa}_{\mu \rho} \Gamma^{\rho}_{\kappa \lambda} -
  \Gamma^{\kappa}_{\kappa \rho}\Gamma^{\rho}_{\mu \lambda}  \bigr)  \,,
  \label{oldG}
\end{align}
where the partial derivatives of the metric in~(\ref{G}) have been rewritten as connection terms. This result is not immediately obvious and the reader's attention is drawn to the indices used in the first quadratic connection term, compared to the index position used in~(\ref{G}).

\subsection{Alternative forms of the bulk term}

The above decomposition of the Ricci scalar is unique in the sense that there is only one canonical boundary term. However, one can write $\overline{\ourG}$ in a variety of different equivalent ways, as was already done in~(\ref{G}) and~(\ref{G2}).

Formulation~(\ref{G2}) turns out to be useful as the field equations will contain the object $\overline{E}_{\mu \nu}{}^{\lambda}$, as will be seen below. Secondly, in the limit of vanishing non-metricity and vanishing torsion, one immediately sees that $-E_{\mu \nu}{}^{\lambda} \partial_{\lambda} g^{\mu \nu}/4 = \ourG$. The first term of~(\ref{G2}) in this limit gives $-\ourG$. This neatly matches~\cite{Boehmer:2021aji} where we wrote $\ourG = \frac{1}{2} M^{\mu \nu}{}_{\lambda} \Gamma_{\mu \nu}{}^{\lambda}$, with $E^{\mu \nu \lambda} = M^{\{\lambda \mu \nu \}}$.

Another previously used form is given by
\begin{align} \label{G3}
  \overline{\mathbf{G}} = g^{\mu \lambda}
  \bigl(\overline{\Gamma}^{\kappa}_{\rho \lambda} \overline{\Gamma}^{\rho}_{\mu \kappa} -
  \overline{\Gamma}^{\kappa}_{\mu \lambda} \overline{\Gamma}^{\rho}_{\rho \kappa} \bigr) +
  \overline{\Gamma}^{\kappa}_{\mu \lambda}  P^{\mu \lambda}{}_{\kappa} \,,
\end{align}
where one has to pay special attention to the index positions in the quadratic part compared to~(\ref{G}) and~(\ref{G3}). This formulation can, for example, be found in~\cite{Hehl:1978a,Hehl:1978b,HehlKerlickHeyde+1976+111+114,HehlKerlickHeyde+1976+524+527,HehlKerlickHeyde+1976+823+827}. Here $P^{\mu \lambda}{}_{\kappa}$ is the Palatini tensor defined by
\begin{align} \label{Palatini}
  P^{\mu \nu}{}_{\lambda} := -Q_{\lambda}{}^{\mu \nu} +  \frac{1}{2} g^{\mu \nu} Q_{\lambda \rho}{}^{\rho} + \delta_{\lambda}^{\mu} Q_{\rho}{}^{\rho \nu} -
  \frac{1}{2} \delta^{\mu}_{\lambda} Q^{\nu}{}_{\rho}{}^{\rho} + T^{\mu}{}_{\lambda}{}^{\nu} + g^{\mu \nu} T^{\rho}{}_{\rho \lambda} +
  \delta^{\mu}_{\lambda} T^{\rho \nu}{}_{\rho} =:
  \frac{\delta \overline{\ourG}}{\delta \overline \Gamma_{\mu \nu}^{\lambda} } \,,
\end{align} 
see Appendix~\ref{A1} for details.
This formulation does have one particular advantage. When varying with respect to the connection, the Palatini tensor will naturally appear in the field equation while in the above it is the object $\overline{E}_{\mu \nu}{}^{\lambda}$ that appears when varying with respect to the metric.

\subsection{Note on boundary terms in teleparallel theories}

Recall from~\cite{Boehmer:2021aji} that the affine Ricci scalar can be decomposed into a sum of various different geometric scalars
\begin{align} \label{fundamental}
\overline{R} = \overbrace{\ourG + \ourB}^{R} + \, T - B_T + Q + B_Q + \ourC \,, 
\end{align}
where $T$ and $Q$ are the torsion and non-metricity scalars, $B_T$ and $B_Q$ are their respective boundary terms, and $\ourC$ are the torsion-non-metricity cross terms. The quantities $\ourG$ and $\ourB$ are the bulk and boundary terms of the Levi-Civita Ricci scalar $R$. One can then show that the affine quantities $\overline{\ourG}$ and $\overline{\ourB}$ are related to the geometric scalars by
\begin{align} \label{G_scalars}
  \overline{\ourG} &= \ourG + T + Q + \ourC \,, \\
  \label{B_scalars}
  \overline{\ourB} &= \ourB - B_T + B_Q \,.
\end{align}

Let us briefly illustrate the above remarks concerning notation and setup in the context of teleparallel gravity. In the teleparallel setting one assumes that the manifold is globally flat, meaning the metric-affine Riemann curvature tensor vanishes $\overline{R}_{\mu \nu \rho}{}^{\lambda} = 0$. Additionally in teleparallel gravity non-metricity is also assumed to vanish $Q_{\lambda \mu \nu} = 0$. These assumptions lead to the following relations between the geometric scalars given in equation~(\ref{fundamental})
\begin{align}
  \ourB - B_T = b_T \,, \qquad \ourG + T = -b_T \,,
\end{align}
where $b_T$ is another boundary term identified in~\cite{Boehmer:2021aji}.
Likewise, for symmetric teleparallel geometries ($\overline{R}_{ \mu \nu \rho}{}^{\lambda}= 0$ and $T^{\lambda \mu \nu}$ = 0) one finds	
\begin{align}
  \ourB + B_Q = b_Q \,, \qquad \ourG + Q = -b_Q \,,
\end{align}
again with $b_Q$ a boundary term~\cite{Boehmer:2021aji}.

An interesting observation is that the requirement for these boundary terms $b_T$ or $b_Q$ to vanish (i.e.~for $\ourG = -T$ or $\ourG = - Q$) is equivalent to requiring $\overline{\ourG}=0$, along with flatness $\overline{R}=0$. This former requirement is a coordinate dependent condition. The \textit{Teleparallel Equivalents of General Relativity}~\cite{Hohmann:2021fpr,BeltranJimenez:2019esp,Nester:1998mp,Aldrovandi:2013wha,Maluf:2013gaa} utilise choosing an affine connection that is flat, such that the Einstein-Hilbert Lagrangian $R$ can be related to the torsion or non-metricity scalars, up to boundary terms. However, from~(\ref{G_scalars}) it is clear that setting just the bulk part to zero $\overline{\ourG}=0$ yields another equivalence between General Relativity and theories with torsion and non-metricity. Clearly this is very much at the heart of the equivalence of the different formulations of General Relativity and emphasises issues like the choice of tetrads in $f(T)$ theories or the choice of coordinates in $f(Q)$ theories. In particular, the condition $b_T = 0$ depends on both frames and coordinates.

\subsection{Variations with respect to metric and connection}

Let us begin with the first order metric-affine Einstein action
\begin{align}
  \label{Einstein_action}
  S[g,\overline{\Gamma}] = \frac{1}{2\kappa} \int \sqrt{-g}\, \overline{\ourG} \, d^4 x \,.
\end{align}
Variations with respect to the metric and the connection lead to 
\begin{align}
  \label{Einstein_var}
  \delta S = \frac{1}{2\kappa} \int \sqrt{-g} \Big[
    \delta g^{\mu \nu} \overline{G}_{\mu \nu} + \delta \overline{\Gamma}^{\lambda}_{\mu \nu} P^{\mu \nu}{}_{\lambda}
    \Big] d^4x = 0 \,,
\end{align}
where $\overline{G}_{\mu \nu}$ is the Einstein tensor of a general affine connection, and $P^{\mu \nu}{}_{\lambda}$ is the Palatini tensor. Details of this calculation can be found in Appendix~\ref{A1}. As is evident from equation~(\ref{Palatini}), this quantity is algebraic in torsion and non-metricity. As a consequence of the projective invariance of the action, the equation is trace-free over the second and third index, implying $P^{\mu \nu}{}_{\nu} = 0$. This tensor therefore has 60 independent components. See Appendix~\ref{appendix_projective} for the projective properties of $P^{\mu \nu}{}_{\lambda}$ as well as the affine quantities previously introduced, which will be useful when studying the modified theory below.
For a more general discussion on projective transformations in metric-affine theories, see~\cite{Afonso:2017bxr}.

In the vacuum or for matter with vanishing hypermomentum, equation~(\ref{Palatini}) can be solved for the connection
\begin{align}
 P^{\mu \nu}{}_{\lambda}= 0 \quad \implies \quad
  \overline{\Gamma}{}^{\lambda}_{\mu \nu} = \Gamma^{\lambda}_{\mu \nu} - \frac{1}{2} Q_{\mu} \delta^{\lambda}_{\nu}
  \quad \mbox{or} \quad
  \overline{\Gamma}{}^{\lambda}_{\mu \nu} = \Gamma^{\lambda}_{\mu \nu}- \frac{1}{3} T_{\mu} \delta^{\lambda}_{\nu} \,,
\end{align}
where $\Gamma$ is the Levi-Civita connection and the additional term is either the non-metricity vector $Q_{\mu} = Q_{\mu}{}^{\rho}{}_{\rho}$ or the torsion vector  $T_{\mu} = T^{\rho}{}_{\rho \mu}$. The 64 equations $P^{\mu \nu}{}_{\lambda} = 0$ fix 60 of the 64 components of $\overline{\Gamma}$, with the 4 free components of $Q_{\mu}$ or $T_{\mu}$ completely unconstrained, again due to the projective invariance. One retrieves GR by assuming either of these vectors to vanish.

Following on from this preliminary discussion, let us now consider the modified first order action
\begin{align}
  S_{\rm{mod}}[g,\overline{\Gamma}] = \frac{1}{2\kappa} \int f(\overline{\ourG}) \sqrt{-g} \, d^4 x \,,
\end{align}
which does not include any boundary terms yet. Variations lead to
\begin{multline}
  \delta S_{\rm{mod}} =\frac{1}{2\kappa} \int \Big[ \delta g^{\mu \nu}
    \Big(
    - \frac{1}{2}g_{\mu \nu} f(\overline{\ourG}) +
    f'(\overline{\ourG})\big( \overline{G}_{\mu \nu} + \frac{1}{2} g_{\mu \nu} \overline{\ourG} \big) +
    \frac{1}{2} f''(\overline{\ourG}) \overline{E}_{\mu \nu}{}^{\lambda} \partial_{\lambda} \overline{\ourG}
    \Big) \\+ \delta \overline{\Gamma}{}^{\lambda}_{\mu \nu}\, P^{\mu \nu}{}_{\lambda} f'(\overline{\ourG})
    \Big] \sqrt{-g}\, d^4 x \,,
\end{multline}
where the object $\overline{E}_{\mu \nu}{}^{\lambda}$ was defined in~(\ref{E}), again see Appendix~\ref{app-a2} for details. This term is absent in General Relativity or other metric-affine theories with $f''(\overline{\ourG}) = 0$, a condition that is often used to define the limit of such theories. It is analogous to the term of the same name used in our previous work~\cite{Boehmer:2021aji}, except with the affine connection being used instead of the Levi-Civita connection.

The connection equation of motion is proportional to Palatini tensor, recall that $\overline{\ourG}$ is also algebraic in torsion and non-metricity. Therefore the connection equation of motion is not dynamical in either torsion or non-metricity, in agreement with the analogous results in the standard Einstein-Palatini variations. For example, it is well-known that torsion does not propagate in Einstein-Cartan theory, which in particular implies the absence of torsional gravitational waves in vacuum. This is in stark contrast to metric perturbations which can propagate through vacuum regions of spacetime.

In order to obtain dynamical behaviour in either of these geometric quantities, we must consider a modified action that includes the boundary term $\overline{\ourB}$. It is precisely those boundary terms which contain the derivatives of torsion and non-metricity. Using the calculus of variations, the required integration by parts introduces additional derivative terms which ultimately give rise to differential as opposed to algebraic equations.

The full modified action, where we allow for an arbitrary function of both the first order term and the boundary term, reads 
\begin{align} \label{action}
  S_{\rm mod}[g, \overline{\Gamma}] =  \frac{1}{2\kappa} \int f(\overline{\ourG}, \overline{\ourB}) \sqrt{-g}\, d^4x \,, 
\end{align}
with the variations leading to
\begin{multline}
  \label{full_variation}
  \delta S_{\rm mod} = \frac{1}{2\kappa} \int \biggl\{
  \delta g^{\mu \nu} \Big[
    - \frac{1}{2} g_{\mu \nu} f +
    f_{,\overline{\ourG}} \big(\overline{G}_{\mu \nu} + \frac{1}{2} g_{\mu \nu} \overline{\ourG}\big) +
    \frac{1}{2} \overline{E}_{\mu \nu}{}^{\lambda} \partial_{\lambda} f_{,\overline{\ourG}} +
    \frac{1}{2} g_{\mu \nu} f_{,\overline{\ourB}} \overline{\ourB} -
    \frac{1}{2}  \overline{E}_{\mu \nu}{}^{\lambda}\partial_{\lambda}f_{,\overline{\ourB}}
    \Big] \\ +
  \delta \overline{\Gamma}{}^{\lambda}_{\mu \nu} \Big[ P^{\mu \nu}{}_{\lambda} f_{,\overline{\ourG}} +
    2 \partial_{\sigma} f_{,\overline{\ourB}}\delta_{\lambda}^{[\mu} g^{\sigma] \nu}
    \Big]
  \biggr\} \sqrt{-g}\, d^4 x \,.
\end{multline}

It is interesting to note that making the Palatini variations of $f(\overline{\ourG})$ and then choosing the Levi-Civita connection (consistent with solving the connection field equation with vanishing hypermomentum) leads back to our metric $f(\ourG)$ gravity theory~\cite{Boehmer:2021aji}. However, performing the same procedure with $f(\overline{\ourG},\overline{\ourB})$ does not yield the metric $f(\ourG,\ourB)$ model. This is perhaps unsurprising as the same situation occurs in Palatini $f(R)$ gravity~\cite{Sotiriou:2006qn,Sotiriou:2008rp,DeFelice:2010aj,Olmo:2011uz}. The reason is clear from the form of~(\ref{affine_EoM}): if one assumes the Levi-Civita connection, the Palatini tensor $P^{\mu \nu}{}_{\lambda}$ vanishes identically. Therefore none of the content of the connection equation of motion of the $f(\ourG)$ model is lost. In contrast, the fourth-order terms coming from the $f(\ldots,\ourB)$ field equation are `lost' in the connection equation of motion. Specifically, the Levi-Civita connection includes derivatives of the metric, which means an integration by parts would be performed on the $\partial_{\lambda} f_{,\ourB}$ term, leading to the fourth-order terms in the Levi-Civita variation of the $f(\ourG,\ourB)$ action.

\subsection{Diffeomorphisms and conservation equations}

To study the modified action~(\ref{action}) under diffeomorphisms, we first look at the changes in the affine quantities $\overline{\ourG}$ and $\overline{\ourB}$ under an infinitesimal coordinate transformation, given by their Lie derivatives. The metric and connection transform under $x^{\mu} \rightarrow \hat{x}^{\mu}(x^{\nu}) = x^{\mu} + \xi^{\mu} $ as
\begin{align}
\delta_{\xi} g_{\mu \nu} = \mathcal{L}_{\xi} g_{\mu \nu} = \xi^{\lambda} \partial_{\lambda} g_{\mu \nu} + \partial_{\mu} \xi^{\lambda} g_{\lambda \nu} + \partial_{\nu} \xi^{\lambda} g_{\lambda \mu} = 2\stackrel{\{ \}}{\nabla}_{(\mu} \xi_{\nu)}
\end{align}
\begin{multline}
\delta_{\xi} \overline{\Gamma}^{\lambda}_{\mu \nu} = \mathcal{L}_{\xi} \overline{\Gamma}^{\lambda}_{\mu \nu} =
\xi^{\rho} \partial_{\rho}  \overline{\Gamma}^{\lambda}_{\mu \nu} - \partial_{\rho} \xi^{\lambda}  \overline{\Gamma}^{\rho}_{\mu \nu} + \partial_{\mu} \xi^{\rho}  \overline{\Gamma}^{\lambda}_{\rho \nu} + \partial_{\nu} \xi^{\rho}  \overline{\Gamma}^{\lambda}_{\mu \rho} + \partial_{\mu} \partial_{\nu} \xi^{\lambda} \\
= \nabla_{\mu} \nabla_{\nu} \xi^{\lambda} + \xi^{\rho} \overline{R}_{\rho \mu \nu}{}^{\lambda} - \nabla_{\mu}(T^{\lambda}{}_{\nu \rho} \xi^{\rho})
\end{multline}
where $\stackrel{\{ \}}{\nabla}$ is the covariant derivative with respect to the Levi-Civita connection, see~\cite{JS1954}. Using these relations, one can straightforwardly calculate 
\begin{align}
\delta_{\xi} \overline{\ourG} &= \mathcal{L}_{\xi} \overline{\ourG} = \xi^{\mu} \partial_{\mu} \overline{\ourG} +\partial_{\mu} \partial_{\nu} \xi^{\lambda} \big(\overline{M}^{\mu \nu}{}_{\lambda} + P^{\mu \nu}{}_{\lambda}\big) \\
\delta_{\xi} \overline{\ourB} &= \mathcal{L}_{\xi} \overline{\ourB} = \xi^{\mu} \partial_{\mu} \overline{\ourB} -\partial_{\mu} \partial_{\nu} \xi^{\lambda} \big(\overline{M}^{\mu \nu}{}_{\lambda} + P^{\mu \nu}{}_{\lambda}\big)
\end{align}
where $\overline{M}$ is defined by the variation of the quadratic part of $\overline{\ourG}$ with respect to the connection
\begin{align}
\overline{M}^{\mu \nu}{}_{\lambda} :=  g^{\rho (\mu} \overline{\Gamma}^{\nu)}_{\lambda \rho} +  g^{\rho (\mu} \overline{\Gamma}^{\nu)}_{\rho \lambda } -  g^{\mu \nu} \overline{\Gamma}^{\rho}_{\rho \lambda }- g^{\rho \sigma} \delta^{(\mu}_{\lambda} \overline{\Gamma}^{\nu)}_{\rho \sigma} \,.
\end{align}
This is the metric-affine equivalent to the Levi-Civita version found in our previous work~\cite{Boehmer:2021aji}. Note that this object is not a tensor. The transformation above can be seen more easily when the bulk term is written in the form~(\ref{G3}), however the calculation is still quite tedious.

Using these transformation properties, an infinitesimal diffeomorphism of the metric-affine Einstein action~(\ref{Einstein_action}) is
\begin{align}
  \label{diffeo1}
  \delta_{\xi} S = \frac{1}{2\kappa} \int \mathcal{L}_{\xi} ( \sqrt{-g}  \,\overline{\ourG}) d^4x  &= \frac{1}{2\kappa} \int  \Big[\partial_{\mu}(\sqrt{-g} \xi^{\mu} \overline{\ourG}) + \partial_{\mu} \partial_{\nu} \xi^{\lambda} \big(\overline{M}^{\mu \nu}{}_{\lambda} + P^{\mu \nu}{}_{\lambda}\big) \sqrt{-g}  \Big] d^4 x \nonumber \\
  &= \textrm{boundary terms} + \frac{1}{2\kappa} \int \xi^{\lambda}  \partial_{\mu} \partial_{\nu} \Big( \sqrt{-g} \big( \overline{M}^{\mu \nu}{}_{\lambda} + P^{\mu \nu}{}_{\lambda} \big) \Big) d^4 x  \, ,
\end{align}
where we have integrated by parts to arrive at the final line. For the unmodified action, the Lagrangian differs from a true coordinate scalar by a boundary term $\overline{\ourG} = \overline{R} - \overline{\ourB}$, and so diffeomorphisms will leave the action invariant $\delta_{\xi} S = 0$, up to boundary terms. Hence, we obtain the identity 
\begin{align} \label{identity1}
 \frac{1}{\sqrt{-g}} \partial_{\mu} \partial_{\nu} \Big( \sqrt{-g} \big( \overline{M}^{\mu \nu}{}_{\lambda} + P^{\mu \nu}{}_{\lambda} \big) \Big) = 0 \,.
\end{align}
This is analogous to the Levi-Civita case~\cite{Boehmer:2021aji}, where performing the same calculation leads to the contracted Bianchi identity, albeit written in a way that is not manifestly covariant. The above identity~(\ref{identity1}) is just the metric-affine version of the contracted Bianchi identity. To see this written in a more conventional form, take the metric and connection variations~(\ref{Einstein_var}) and let this be generated by the infinitesimal transformation $\xi$, which leads to
\begin{align} \label{diffeo2}
  \delta_{\xi} S &= \frac{1}{2\kappa} \int \sqrt{-g} \Big[
    \delta_{\xi} g^{\mu \nu} \overline{G}_{\mu \nu} + \delta_{\xi} \overline{\Gamma}{}^{\lambda}_{\mu \nu} P^{\mu \nu}{}_{\lambda}
    \Big] d^4x \nonumber \\
    &= \frac{1}{2\kappa} \int \sqrt{-g} \Big[ -2\stackrel{\{ \}}{\nabla}_{(\mu}  \xi_{\nu)} \overline{G}{}^{\mu \nu} + \big
    ( \nabla_{\mu} \nabla_{\nu} \xi^{\lambda} + \xi^{\rho} \overline{R}_{\rho \mu \nu}{}^{\lambda} - \nabla_{\mu}(T^{\lambda}{}_{\nu \rho } \xi^{\rho}) \big) P^{\mu \nu}{}_{\lambda} \Big] d^4 x  \nonumber \\
    &=  \textrm{boundary terms} + \frac{1}{2\kappa} \int \sqrt{-g} \xi^{\mu} \Big[ 2\stackrel{\{ \}}{\nabla}_{\nu} \overline{G}{}^{(\nu}{}_{\mu)}  \nonumber \\
    & + \frac{1}{\sqrt{-g}}\nabla_{\nu} \nabla_{\lambda} (\sqrt{-g} P^{\lambda \nu}{}_{\mu})
    + \frac{1}{\sqrt{-g}}  \nabla_{\nu}(\sqrt{-g} P^{\nu \lambda}{}_{\rho}) T^{\rho}{}_{\lambda \mu} + \overline{R}_{\mu \nu \lambda}{}^{\rho} P^{\nu \lambda}{}_{\rho} \Big] d^4 x = 0 \,.
\end{align}

Discarding boundary terms, we are led to the identity 
\begin{align} \label{identity2}
2 \stackrel{\{ \}}{\nabla}_{\nu}  \overline{G}{}^{(\nu}{}_{\mu )}   
    + \frac{1}{\sqrt{-g}}\nabla_{\nu} \nabla_{\lambda} (\sqrt{-g} P^{\lambda \nu}{}_{\mu})
    + \frac{1}{\sqrt{-g}}  \nabla_{\nu}(\sqrt{-g} P^{\nu \lambda}{}_{\rho}) T^{\rho}{}_{\lambda \mu} + \overline{R}_{\mu \nu \lambda}{}^{\rho} P^{\nu \lambda}{}_{\rho} \equiv 0 \,,
\end{align}
which is manifestly covariant. Both forms~(\ref{identity1}) and~(\ref{identity2}) are equivalent.

Next let us study the modified, non-covariant action. One could take the same route as above, replacing the metric and connection field equations in equation~(\ref{diffeo2}) with their modified counterparts, given by~(\ref{field1}) and~(\ref{field2}) respectively. This is how one obtains the conservation laws in for example $f(R)$ gravity and Palatini $f(\overline{R})$ gravity~\cite{1993GReGr..25..461H,Koivisto:2005yk}. Instead, we will take the former approach, as in equation~(\ref{diffeo1}), where we obtain a conservation law from the infinitesimal transformations of $\overline{\ourG}$ and $\overline{\ourB}$ that takes a more compact form. Under an infinitesimal diffeomorphism the modified action transforms as
\begin{align} 
\delta_{\xi} S_{\textrm{mod}} &= \frac{1}{2\kappa} \int \mathcal{L}_{\xi} ( \sqrt{-g}  f(\overline{\ourG},\overline{\ourB}) ) d^4x 
\\
& \begin{multlined} = \textrm{boundary terms}  \\
+ \frac{1}{2\kappa} \int \xi^{\lambda}  \partial_{\mu} \partial_{\nu} \Big[ \sqrt{-g} \big( \overline{M}^{\mu \nu}{}_{\lambda} + P^{\mu \nu}{}_{\lambda} \big) \Big( \frac{\partial f(\overline{\ourG}, \overline{\ourB})}{\partial \overline{\ourG}}  -  \frac{\partial f(\overline{\ourG}, \overline{\ourB})}{\partial \overline{\ourB}} \Big)  \Big] d^4 x  \,,
\end{multlined}
\end{align}
which for arbitrary $\xi$ leaves the expression
\begin{align}
\frac{1}{\sqrt{-g}} \partial_{\mu} \partial_{\nu} \Big[ \sqrt{-g} \big( \overline{M}^{\mu \nu}{}_{\lambda} + P^{\mu \nu}{}_{\lambda} \big) \big( f_{,\overline{\ourG}}  -  f_{,\overline{\ourB}} \big)  \Big] \,.
\end{align}
Unlike for an invariant scalar action, the above transformation need not vanish identically. The expression is also non-covariant and depends on the choice of coordinates, due to the non-tensorial nature of the bulk and boundary terms. One immediately sees that for the choice of function $f(\overline{\ourG} + \overline{\ourB})=f(\overline{R})$ the expression vanishes, reflecting the diffeomorphism invariance of the Palatini $f(\overline{R})$ theories.

If we insist that our gravitational action remain invariant under the above transformations, we obtain the conservation law
\begin{align} \label{cons1}
 \partial_{\mu} \partial_{\nu} \Big[ \sqrt{-g} \big( \overline{M}^{\mu \nu}{}_{\lambda} + P^{\mu \nu}{}_{\lambda} \big) \big( f_{,\overline{\ourG}}  -  f_{,\overline{\ourB}} \big)  \Big] = 0 \,,
\end{align}
which depends on both the geometry $g_{\mu \nu}$ and the choice of coordinates $x^{\mu}$. If we also include minimally coupled matter in our total action, we have the following on-shell conservation law
\begin{multline} \label{cons2}
\frac{1}{2\kappa}\partial_{\mu} \partial_{\nu} \Big[ \sqrt{-g} \big( \overline{M}^{\mu \nu}{}_{\lambda} + P^{\mu \nu}{}_{\lambda} \big) \big( f_{,\overline{\ourG}}  -  f_{,\overline{\ourB}} \big)  \Big]  \\
-\sqrt{-g} \stackrel{\{ \}}{\nabla}_{\mu} T^{\mu}{}_{\lambda} - \nabla_{\nu} \nabla_{\mu}(\sqrt{-g} \Delta^{\mu \nu}{}_{\lambda}) - \sqrt{-g} \overline{R}_{\lambda \mu \nu}{}^{\rho} \Delta^{\mu \nu}{}_{\rho} - \nabla_{\mu}(\sqrt{-g} \Delta^{\mu \nu}{}_{\rho}) T^{\rho}{}_{\nu \lambda}
  = 0 \,.
\end{multline}
For a consistent variation principle of the total action $S_{\rm{tot}} = S_{\rm{mod}} + S_{\rm{matter}}$ we require the above equation~(\ref{cons2}) to be satisfied. This is the most general case, as we do not consider non-minimal couplings here. If the matter action $S_{\rm{matter}}$ is a coordinate scalar then we must impose that the first of these equations~(\ref{cons1}) is satisfied. In these cases, one obtains the usual metric-affine matter-hypermomentum conservation laws
\begin{align} \label{cons3}
\sqrt{-g} \stackrel{\{ \}}{\nabla}_{\mu} T^{\mu}{}_{\lambda} + \nabla_{\nu} \nabla_{\mu}(\sqrt{-g} \Delta^{\mu \nu}{}_{\lambda}) + \sqrt{-g} \overline{R}_{\lambda \mu \nu}{}^{\rho} \Delta^{\mu \nu}{}_{\rho} + \nabla_{\mu}(\sqrt{-g} \Delta^{\mu \nu}{}_{\rho}) T^{\rho}{}_{\nu \lambda}
  = 0 \,,
\end{align}
see for example~\cite{Hohmann:2021fpr}. Vanishing hypermomentum leads to the standard energy-momentum conservation law. 

More generally, one could conceive scenarios where the matter action is also non-covariant, and it is only the total action that remains invariant under diffeomorphisms. Some of these possibilities are discussed in more detail for the Levi-Civita case in~\cite{Boehmer:2021aji}. This could be implemented by allowing for non-minimal couplings between the pseudo-scalars $\overline{\ourG}$ and $\overline{\ourB}$ and matter, in which case~(\ref{cons2}) would need to be modified.

Lastly, relating to the non-covariance of these theories, we note the interesting possibility of restoring the full diffeomorphism invariance of the theory by using the St{\"u}ckelberg trick~\cite{Gao:2014soa}. In this sense, the theory can be thought of as a gauge-fixed version of a fully covariant theory, where extra degrees of freedom should become apparent. This is exactly the case for the Levi-Civita version of $f(\ourG)$ gravity, which turns out to give rise to the modified symmetric teleparallel theories $f(Q)$ gravity when covariance is restored~\cite{BeltranJimenez:2022azb}. There, the extra degrees of freedom can be attributed to the independent components of the symmetric teleparallel connection, the coordinate functions often denoted by $\xi(x)$. These are precisely the St{\"u}ckelberg fields, which trivialise in the coincident gauge $\xi^{\mu}(x)=x^{\mu}$ where $f(Q)$ reduces to $f(\ourG)$, see~\cite{BeltranJimenez:2017tkd,BeltranJimenez:2022azb,Boehmer:2021aji}. This is also reminiscent of the diffeomorphism-invariant formulation of classical Unimodular gravity which introduces St{\"u}ckelberg fields to restore the full symmetry of the theory~\cite{Kuchar:1991xd,Padilla:2014yea}.

This would be an interesting avenue to explore in the future, as it is not obvious what type of theory this procedure would lead to in this metric-affine case, nor how these extra degrees of freedom would manifest. For the remainder of this work we will assume that~(\ref{cons1}) is satisfied by using an appropriate choice of coordinates. 

\section{New models and relations to other theories}

\subsection{General field equations with source terms}

The field equations~(\ref{full_variation}) together with appropriate source terms give rise to the full `metric' field equation
\begin{align}
  \label{field1}
  - \frac{1}{2} g_{\mu \nu} f +
  f_{,\overline{\ourG}} \big(\overline{G}_{(\mu \nu)} + \frac{1}{2} g_{\mu \nu} \overline{\ourG}\big) +
  \frac{1}{2} \overline{E}_{(\mu \nu)}{}^{\lambda} \partial_{\lambda} f_{,\overline{\ourG}} +
  \frac{1}{2} g_{\mu \nu} f_{,\overline{\ourB}} \overline{\ourB} -
  \frac{1}{2} \overline{E}_{(\mu \nu)}{}^{\lambda} \partial_{\lambda}f_{,\overline{\ourB}} =
  \kappa {}^{(\overline{\Gamma})} T_{\mu \nu} \,,
\end{align}
and the corresponding `connection' field equation
\begin{align}
  \label{field2}
  P^{\mu \nu}{}_{\lambda} f_{,\overline{\ourG}} +
  2 \partial_{\rho} f_{,\overline{\ourB}} \delta_{\lambda}^{[\mu} g^{\rho] \nu} =
  2\kappa \Delta^{\mu \nu}{}_{\lambda} \,.
\end{align}
Here the matter action is given by
\begin{align}
  S_{\rm matter} = S_{\rm matter}[g,\overline{\Gamma},\phi] =
  \int L_{\rm matter}(g,\overline{\Gamma},\phi) \sqrt{-g}\, d^4x \,,
\end{align}
and is assumed to depend on the metric and an independent connection. This is a crucial point for what follows. When working on manifolds with vanishing non-metricity, variations of the metric and connection can no longer be assumed to be independent. Consequently, variations with respect to the metric, in such cases, will contain additional terms as the metric's first derivatives enter the connection. This can be dealt with in several different ways, from our point of view the most appropriate approach is via the use of Lagrange multipliers. We will elaborate on this as and when needed.

The symmetric (metrical) energy-momentum tensor is defined in analogy to the Hilbert energy-momentum tensor
\begin{align}
  {}^{(\overline{\Gamma})} T_{\mu \nu} := -\frac{2}{\sqrt{-g}}\frac{\delta (\sqrt{-g} L_{\rm matter})}{\delta g^{\mu \nu}} \,.
\end{align}
The superscript $\overline{\Gamma}$ here makes it explicit that in the matter action the metric and connection are treated as completely independent variables, see also~\cite{Hehl:1978a} for discussions relating to the definition of the energy-momentum tensor depending on the choice of variables. A modern treatment of this issue can also be found in~\cite{Hehl:1994ue}.

The variation with respect to the connection is defined as the hypermomentum
\begin{align}
  \Delta^{\mu \nu}{}_{\lambda} := -\frac{1}{\sqrt{-g}}\frac{\delta (\sqrt{-g} L_{\rm matter})}{\delta \overline{\Gamma}_{\mu \nu}^{\lambda}} \,,
\end{align}
see for instance~\cite{HehlKerlickHeyde+1976+111+114,HehlKerlickHeyde+1976+524+527,HehlKerlickHeyde+1976+823+827,Obukhov:1993pt,Babourova:1998mgh} and~\cite{Iosifidis:2020gth,Iosifidis:2020upr,Hohmann:2021ast} for some more recent references with application in cosmology.

\subsection{General Relativity and its metric-affine generalisations}
\label{subsec:metricaffineGR}

Let us start by looking at functions linear in $\overline{\ourG}$, which will include metric-affine gravity models like Einstein-Cartan theory or General Relativity in the simplest of cases. Setting $f(\overline{\ourG},\overline{\ourB}) = \ourG$, the field equations now reduce to
\begin{align}
  \overline{G}_{(\mu \nu)} &= \kappa\, {}^{(\overline{\Gamma})}T_{\mu \nu} \,, \\
  P^{\mu \nu}{}_{\lambda} &= 2\kappa \Delta^{\mu \nu}{}_{\lambda} \,,
\end{align}
and can be found, for example, in~\cite{Hehl:1978b}. Due to the projective invariance of $P^{\mu \nu}{}_{\lambda}$, the second field equation is only consistent if the matter Lagrangian is such that $\Delta^{\mu \nu}{}_{\nu} = 0$.  

Einstein-Cartan theory~\cite{FH1976} follows from the above when setting non-metricity to zero. However, as explained, this procedure cannot be directly put into these equations as the metric and connection variations are no longer fully independent and one would arrive at incorrect field equations. Let us supplement our action by the following Lagrange multiplier
\begin{align}
  S_Q = -\int \frac{1}{2 }\lambda^{\mu \nu \rho} Q_{\mu \nu \rho} \sqrt{-g}\, d^4 x =
  \int \frac{1}{2 }\lambda^{\mu \nu \rho} \nabla_{\mu} g_{\nu \rho} \sqrt{-g}\, d^4 x\,,
\end{align}
which ensures that non-metricity vanishes, the chosen sign is a matter of conventions. We have $\lambda^{\mu \nu \lambda} = \lambda^{\mu \lambda \nu}$ to match the symmetry properties of non-metricity. We now choose as independent variables $\{g,\overline{\Gamma},\lambda\}$ and any matter fields present. Writing out this covariant derivative explicitly and using integration by parts, we can write the constraint as
\begin{align}
  S_Q = \frac{1}{2 }\int \Bigl[
    -g_{\nu \rho} \frac{1}{\sqrt{-g}}\partial_{\mu} (\sqrt{-g}\lambda^{\mu \nu \rho})
    -\lambda^{\mu \nu \sigma} \overline{\Gamma}^{\rho}_{\mu \nu} g_{\rho \sigma }-\lambda^{\mu \nu \sigma} \overline{\Gamma}_{\mu \sigma}^{\rho} g_{\nu \rho}
    \Bigr] \sqrt{-g}\, d^4 x\,,
\end{align}
from which the metric and connection variations can be straightforwardly computed.
Our complete metric field equation now reads
\begin{multline}
  \label{fieldc1}
  - \frac{1}{2} g_{\mu \nu} f +
  f_{,\overline{\ourG}} \big(\overline{G}_{(\mu \nu)} + \frac{1}{2} g_{\mu \nu} \overline{\ourG}\big) + 
  \frac{1}{2} \overline{E}_{(\mu \nu)}{}^{\sigma} \partial_{\sigma} f_{,\overline{\ourG}} \\ +
  \frac{1}{2} g_{\mu \nu} f_{,\overline{\ourB}} \overline{\ourB} -
  \frac{1}{2} \overline{E}_{(\mu \nu)}{}^{\sigma} \partial_{\sigma}f_{,\overline{\ourB}} +
  \kappa \nabla_{\sigma} \lambda^{\sigma}{}_{\mu \nu} + \kappa T^{\rho}{}_{\sigma \rho}\lambda^{\sigma}{}_{\mu \nu} =
  \kappa {}^{(\overline{\Gamma})}T_{\mu \nu} \,,
\end{multline}
while the connection and constraint equation are given by
\begin{align}
  \label{fieldc2}
  P^{\mu \nu}{}_{\rho} f_{,\overline{\ourG}} +
  2 \partial_{\sigma} f_{,\overline{\ourB}} \delta_{\rho}^{[\mu} g^{\sigma] \nu} - 2 \kappa \lambda^{\mu \nu}{}_{\rho}
  &= 2\kappa \Delta^{\mu \nu}{}_{\rho} \,, \\
  \label{fieldc3}
  \nabla_{\mu} g_{\nu \rho} &= 0 \,.
\end{align}
Note that we have already implemented the constraint equation to remove any non-metricity terms which would otherwise be present in the field equations.

As before, setting $f(\overline{\ourG},\overline{\ourB}) = \overline{\ourG}$, the field equations reduce to
\begin{align}
  \label{fieldd1}
  \overline{G}_{(\mu \nu)} + \kappa (\nabla_{\rho} + T^{\sigma}{}_{\rho \sigma}) \lambda^{\rho}{}_{\mu \nu} &=
  \kappa {}^{(\overline{\Gamma})}T_{\mu \nu} \,, \\
  \label{fieldd2}
  P_{\mu \nu \rho} - 2\kappa \lambda_{\mu \nu \rho} &= 2\kappa \Delta_{\mu \nu \rho} \,.
\end{align}
One should now pay particular attention to the second field equation which is algebraic in all quantities. Using $Q=0$ in~(\ref{Palatini}) one finds
\begin{align}
  \label{M2}
  P_{\mu \nu \rho} = T_{\mu \rho \nu} - g_{\mu \nu} T^{\sigma}{}_{\rho \sigma} + g_{\mu \rho} T^{\sigma}{}_{\nu \sigma} \,,
\end{align}
which immediately implies $P_{\mu(\nu \rho)}=0$ as this object is skew-symmetric over the last two indices. Hence, the symmetric part of~(\ref{fieldd2}) yields the Lagrange multiplier to be $\lambda_{\mu \nu \rho} = - \Delta_{\mu(\nu \rho)}$, whilst the skew-symmetric part of this equation gives $P_{\mu \nu \rho} = 2\kappa \Delta_{\mu[\nu \rho]}$ as expected from Einstein-Cartan theory. Lastly, substitution of these results back into the first field equation~(\ref{fieldd1}) gives the desired field equations of Einstein-Cartan theory
\begin{align}
  \label{fieldd1b}
  \overline{G}_{(\mu \nu)} &=
  \kappa {}^{(\overline{\Gamma})} T_{\mu \nu} + \kappa (\nabla_{\rho} + T^{\sigma}{}_{\rho \sigma}) \Delta^{\rho}{}_{(\mu \nu)}\,, \\
  \label{fieldd1c}
  P_{\mu \nu \rho} &= 2\kappa \Delta_{\mu[\nu \rho]} \,.
\end{align}

Let us make a brief comment regarding the need to use Lagrange multipliers in the metric-affine setting. We found explicitly that $\lambda_{\mu \nu \rho} = - \Delta_{\mu(\nu \rho)}$. For arbitrary matter sources $\Delta_{\mu(\nu \rho)} \neq 0$ in general so that $\lambda_{\mu \nu \rho} \neq 0$ in general. Consequently, one cannot simply set non-metricity to zero after performing the variations since one would lose the second term on the right-hand side of~(\ref{fieldd1b}).

In the case of vanishing non-metricity $P^{\mu \nu}{}_{\lambda}$ simplifies to
\begin{align}
  \label{Mqzero}
  P^{\mu \nu}{}_{\lambda} = T^{\mu}{}_{\lambda}{}^{\nu} + g^{\mu \nu} T^{\rho}{}_{\rho \lambda} + \delta^{\mu}_{\lambda} T^{\rho \nu}{}_{\rho} \,,
\end{align}
which is the familiar left-hand side of the `torsional' field equation. Sometimes the object appearing on the left-hand side is called the modified torsion tensor. Closely related to this object is the so-called superpotential of teleparallel gravity, see for instance~\cite{RA2013}. Moreover, in the absence of non-metricity the corresponding source term is known as the spin angular momentum tensor, in which case one often uses the notation $\tau_{\mu \nu \lambda} := \Delta_{\mu[\nu \lambda]}$.

\subsection{Palatini \texorpdfstring{$f(R)$}{f(R)} gravity}

Palatini $f(\overline{R})$ gravity is also included in our approach and achieved by simply setting $f(\overline{\ourG},\overline{\ourB}) = f(\overline{\ourG} + \overline{\ourB})$. This follows from the basic decomposition of the Ricci scalar that underlies this work. Here we work in the general affine setting with torsion and non-metricity present.

For the metric field equations, the $\overline{E}$ terms cancel  
\begin{align}
  - \frac{1}{2} g_{\mu \nu} f + f'(R) \bigl(\overline{G}_{(\mu \nu)} +
  \frac{1}{2} g_{\mu \nu} \overline{\ourG} + \frac{1}{2} g_{\mu \nu} \overline{\ourB} \bigr) =
  \kappa  {}^{(\overline{\Gamma})} T_{\mu \nu} \,,
\end{align}
which we now write as follows
\begin{align} \label{f(R)}
  f'(\overline{R}) \overline{R}_{(\mu \nu)} - \frac{1}{2} g_{\mu \nu} f(\overline{R}) =
  \kappa  {}^{(\overline{\Gamma})} T_{\mu \nu} \,,
\end{align}
where we used the relations $\overline{G}_{\mu \nu} + g_{\mu \nu} \overline{R}/2 = \overline{R}_{\mu \nu}$ and $\overline{\ourG}+\overline{\ourB} = \overline{R}$. The connection equation is 
\begin{align}
  \label{f(R)_connection}
  P^{\mu \nu}{}_{\lambda} f'(R) + 2 \partial_{\rho} f'(\overline{R}) \delta_{\lambda}^{[\mu} g^{\rho] \nu} =
  2\kappa \Delta^{\mu \nu}{}_{\lambda} \,.
\end{align}
Let us unpack the connection equation and work it into a more familiar form by first noting that
\begin{multline}
  \label{covar_deriv}
  \frac{2}{\sqrt{-g}} \nabla_{\rho} (f'(R) \sqrt{-g}  \delta_{\lambda}^{[\mu} g^{\rho] \nu}) =
  2 \partial_{\rho} f'(\overline{R}) \delta_{\lambda}^{[\mu} g^{\rho] \nu} \\
  + f'(\overline{R})
  \Big(\frac{1}{2} Q_{\lambda}{}^{\rho}{}_{\rho} g^{\mu \nu} -
  \frac{1}{2} Q^{\nu \rho}{}_{\rho} \delta^{\mu}_{\lambda} +
  Q_{\rho}{}^{\rho \nu} \delta^{\mu}_{\lambda} - Q_{\lambda}{}^{\mu \nu }\Big)
  \\ =
  2 \partial_{\rho} f'(\overline{R}) \delta_{\lambda}^{[\mu} g^{\rho] \nu}  +
  f'(\overline{R})  P^{\mu \nu}{}_{\lambda} -  f'(\overline{R})(T^{\mu}{}_{\lambda}{}^{\nu} +
  g^{\mu \nu} T^{\rho}{}_{\rho \lambda} + \delta^{\mu}_{\lambda} T^{\rho \nu}{}_{\rho}) \,,
\end{multline}
where in the first line we expanded out the covariant derivatives of the metric in terms of the non-metricity tensor, and in the second line we inserted the definition of $P^{\mu \nu}{}_{\lambda}$. Putting together~(\ref{f(R)_connection}) and~(\ref{covar_deriv}) we arrive at
\begin{align}
  \label{f(R)_connection_final}
  \frac{2}{\sqrt{-g}} \nabla_{\rho} (f'(R) \sqrt{-g} \delta_{\lambda}^{[\mu} g^{\rho] \nu}) +
  f'(\overline{R})(T^{\mu}{}_{\lambda}{}^{\nu} + g^{\mu \nu} T^{\rho}{}_{\rho \lambda} +
  \delta^{\mu}_{\lambda} T^{\rho \nu}{}_{\rho} )  =2 \kappa \Delta^{\mu \nu}{}_{\lambda} \,.
\end{align}
This is the familiar form of the $f(\overline{R})$ connection equation of motion, as reported for example in~\cite{Sotiriou:2006qn}.

Taking the trace of the metric equation~(\ref{f(R)}) leads to the well known algebraic  relation $-2f(\overline{R}) + f'(\overline{R})\overline{R} = \kappa T$. This implies an algebraic relation between the Ricci scalar $\overline{R}$ and the trace of the energy-momentum tensor $T$. Contrast this with the $f(\overline{\ourG},\overline{\ourB})$ equations~(\ref{field1}), where the trace leads to a dynamical equation as opposed to an algebraic one. This leads to a number of interesting possibilities beyond what can be found in the metric-affine $f(\overline{R})$ theories.

\subsection{A modified Einstein-Cartan theory}

Let us now return to the field equations~(\ref{fieldc1})--(\ref{fieldc3}) and assume $f(\overline{\ourG},\overline{\ourB}) = f(\overline{\ourG})$ so that $f_{,\overline{\ourB}} = 0$. Moreover, we assume that non-metricity vanishes via the introduction of the Lagrange multiplier, in which case the field equations simplify to
\begin{align}
  \label{fieldd1EC}
  - \frac{1}{2} g_{\mu \nu} f +
  f_{,\overline{\ourG}} \big(\overline{G}_{(\mu \nu)} + \frac{1}{2} g_{\mu \nu} \overline{\ourG}\big) +
  \frac{1}{2} \overline{E}_{(\mu \nu)}{}^{\rho} \partial_{\rho} f_{,\overline{\ourG}} +
  \kappa \nabla_{\rho} \lambda^{\rho}{}_{\mu \nu} + \kappa T^{\sigma}{}_{\rho \sigma}\lambda^{\rho}{}_{\mu \nu} &=
  \kappa {}^{(\overline{\Gamma})}T_{\mu \nu} \,, \\
  P^{\mu \nu}{}_{\rho} f_{,\overline{\ourG}} - 2 \kappa \lambda^{\mu \nu}{}_{\rho} &= 2\kappa \Delta^{\mu \nu}{}_{\rho} \,.
\end{align}
As before, the Palatini tensor is now skew-symmetric in the last pair of indices and we find $-\lambda_{\mu \nu \rho} = \Delta_{\mu(\nu \rho)}$ and $P_{\mu \nu \rho} f_{,\overline{\ourG}} = 2\kappa \Delta_{\mu[\nu \rho]}$. Hence, we can write the first and second field equations in the form
\begin{align}
  - \frac{1}{2} g_{\mu \nu} f +
  f_{,\overline{\ourG}} \big(\overline{G}_{(\mu \nu)} + \frac{1}{2} g_{\mu \nu} \overline{\ourG}\big) +
  \frac{1}{2} \overline{E}_{(\mu \nu)}{}^{\rho} \partial_{\rho} f_{,\overline{\ourG}} &=
  \kappa \Bigl({}^{(\overline{\Gamma})}T_{\mu \nu} + (\nabla_{\rho} + T^{\sigma}{}_{\rho \sigma})\Delta^{\rho}{}_{( \nu \mu)}\Bigr)\,,
  \label{fieldd2EC} \\
  P_{\mu \nu \rho} f_{,\overline{\ourG}} &= 2\kappa \Delta_{\mu[\nu \rho]} \,.
  \label{fieldd2b}
\end{align}
The right-hand side of equation~(\ref{fieldd2EC}) takes the familiar form of the symmetrised canonical energy-momentum tensor. This is the $f(\overline{\ourG})$ modified Einstein-Cartan theory field equation.

It is well known that in Einstein-Cartan theory one can remove the symmetry brackets of the first field equation and consider the full equation instead. It then turns out that the skew-symmetric part of this equation coincides with the second field equation. If we follow this approach here and consider the skew-symmetric part of~(\ref{fieldd2EC}), we find
\begin{align}
  f_{,\overline{\ourG}} \overline{G}_{[\mu \nu]} +
  \frac{1}{2} \overline{E}_{[\mu \nu]}{}^{\rho} \partial_{\rho} f_{,\overline{\ourG}} =
  \kappa (\nabla_{\rho} + T^{\sigma}{}_{\rho \sigma})\Delta^{\rho}{}_{[\mu \nu]} =
  \frac{1}{2} (\nabla_{\rho} + T^{\sigma}{}_{\rho \sigma})(P^{\rho}{}_{ \nu \mu } f_{,\overline{\ourG}}) \,.
  \label{fieldd3}
\end{align}
After multiplying out the right-hand side using the product rule, and taking into account the geometrical identify $\overline{G}_{[\mu \nu]} = (\nabla_{\rho} + T^{\sigma}{}_{\rho \sigma})P^{\rho}{}_{ \nu \mu }/2$ we arrive at the interesting equation
\begin{align}
  \frac{1}{2} \overline{E}_{[\mu \nu]}{}^{\rho} \partial_{\rho} f_{,\overline{\ourG}} =
  \frac{1}{2} P^{\rho}{}_{ \nu \mu} \partial_{\rho} f_{,\overline{\ourG}} \,,
  \label{fieldd4}
\end{align}
which immediately gives $\overline{E}_{[\mu \nu]}{}^{\rho} = P^{\rho}{}_{ \nu \mu}$. This equation can be seen as determining the skew-symmetric part of $\overline{E}_{[\mu \nu]}{}^{\rho}$ such that, in complete analogy with Einstein-Cartan theory, the full first field equation contains the second field equation. This allows us the define a new object as follows
\begin{align}
  \mathcal{E}_{\mu \nu}{}^{\rho} &:= \overline{E}_{(\mu \nu)}{}^{\rho} + \overline{E}_{[\mu \nu]}{}^{\rho}
  \nonumber \\ &:=
  2 \overline{\Gamma}^{\rho}_{(\mu \nu)} -
  2 \delta^{\rho}_{(\mu} \overline{\Gamma}^{\lambda}_{|\lambda |\nu)} - g_{\mu \nu} (g^{\kappa \lambda} \overline{\Gamma}^{\rho}_{\lambda \kappa} -
  g^{\kappa \rho} \overline{\Gamma}^{\lambda}_{\lambda \kappa}) + P^{\rho}{}_{ { \nu \mu} } \nonumber \\
  &= 2 \overline{\Gamma}{}^{\rho}_{ { \mu \nu} } + 2 \delta^{\rho}_{[ {  \mu} } \overline{\Gamma}^{\lambda}_{{\nu} ]\lambda} - 2 \delta^{\rho}_{ { \mu } } \overline{\Gamma}{}^{\lambda}_{\lambda { \nu} } - g_{\mu \nu} (g^{\kappa \lambda} \overline{\Gamma}^{\rho}_{\lambda \kappa} -
  g^{\kappa \rho} \overline{\Gamma}^{\lambda}_{\lambda \kappa})  \,.
  \label{fieldd5}
\end{align}
We are now able to re-write the complete first field equation
\begin{align}
  - \frac{1}{2} g_{\mu \nu} f +
  f_{,\overline{\ourG}} \big(\overline{G}_{\mu \nu} + \frac{1}{2} g_{\mu \nu} \overline{\ourG}\big) +
  \frac{1}{2} \mathcal{E}_{\mu \nu}{}^{\lambda} \partial_{\lambda} f_{,\overline{\ourG}} =
  \kappa \Bigl({}^{(\overline{\Gamma})}T_{\mu \nu} + (\nabla_{\lambda} + T^{\sigma}{}_{\lambda \sigma})\Delta^{\lambda}{}_{{ \nu \mu } }\Bigr)\,,
  \label{fieldd6}
\end{align}
so that its skew-symmetric part coincides with the second field equation. This result is somewhat surprising but illustrates the intricate interplay between the two field equations and the remaining freedom one could explore in this setting.

\subsection{A modified Einstein-Cartan theory with boundary terms}

Let us now work with the fully general $f(\overline{\ourG},\overline{\ourB})$ theory, given by equations~(\ref{fieldc1})-(\ref{fieldc3}). The connection equation with lowered indices is
\begin{align}
  P_{\mu \nu \rho} f_{,\overline{\ourG}} + 2 g_{\mu[\rho} \partial_{\nu]} f_{,\overline{\ourB}} - 2 \kappa \lambda_{\mu \nu \rho} = 2\kappa \Delta_{\mu \nu \rho} \,,
  \label{ECmod1}
\end{align}
from which we immediately see the symmetric part $\lambda_{\mu \nu \rho} = - \Delta_{\mu(\nu \rho)}$ is the same as before. The skew-symmetric part now includes an additional dynamical term
\begin{align}
  P_{\mu \nu \rho} f_{,\overline{\ourG}} + 2 g_{\mu[\rho} \partial_{\nu]} f_{,\overline{\ourB}} = 2\kappa \Delta_{\mu[\nu \rho]}\,.
  \label{ECmod2}
\end{align}
The metric field equation is then
\begin{multline}
  - \frac{1}{2} g_{\mu \nu} f +
  f_{,\overline{\ourG}} \big(\overline{G}_{(\mu \nu)} + \frac{1}{2} g_{\mu \nu} \overline{\ourG}\big) +
  \frac{1}{2} \overline{E}_{(\mu \nu)}{}^{\lambda} \partial_{\lambda} f_{,\overline{\ourG}} \\ +
  \frac{1}{2} g_{\mu \nu} f_{,\overline{\ourB}} \overline{\ourB} -
  \frac{1}{2} \overline{E}_{(\mu \nu)}{}^{\lambda} \partial_{\lambda}f_{,\overline{\ourB}} =
  \kappa \Big( {}^{(\overline{\Gamma})}T_{\mu \nu} + (\nabla_{\lambda} + T^{\sigma}{}_{\lambda \sigma}) \Delta^{\lambda}{}_{({\nu \mu})} \Big) \,.
  \label{ECmod3}
\end{multline}
Again, just as before, we now remove the symmetry brackets and consider the full field equation, first looking at the skew-symmetric part
\begin{align}
  f_{,\overline{\ourG}} \overline{G}_{[\mu \nu]} +
  \frac{1}{2} \overline{E}_{[\mu \nu]}{}^{\lambda} \partial_{\lambda} f_{,\overline{\ourG}} -
  \frac{1}{2} \overline{E}_{[\mu \nu]}{}^{\lambda} \partial_{\lambda} f_{,\overline{\ourB}} =
  \frac{1}{2} (\nabla_{\lambda} + T^{\sigma}{}_{\lambda \sigma})  \Big( P^{\lambda}{}_{ { \nu \mu} } f_{,\overline{\ourG}} + 2 \delta^{\lambda}_{[ {\mu} } \partial_{ { \nu} ]} f_{,\overline{\ourB}}  \Big)\,,
  \label{ECmod4}
 \end{align}
where on the right hand side we have used the skew part of the connection equation. Using the geometric identity $\overline{G}_{[\mu \nu]} = (\nabla_{\lambda} + T^{\sigma}{}_{\lambda \sigma})P^{\lambda}{}_{{ \nu \mu } }/2$ we find another interesting equation
\begin{align}
  \label{fGB1}
  \frac{1}{2}  \partial_{\lambda} f_{,\overline{\ourG}}  \Big( \overline{E}_{[\mu \nu]}{}^{\lambda} - P^{\lambda}{}_{ { \nu \mu} } \Big) - \frac{1}{2}\overline{E}_{[\mu \nu]}{}^{\lambda} \partial_{\lambda} f_{,\overline{\ourB}} = (\nabla_{\lambda} + T^{\sigma}{}_{\lambda \sigma}) \delta^{\lambda}_{[{ \mu} } \partial_{ { \nu}]}  f_{,\overline{\ourB}} \,.
\end{align} 
The part in the first bracket was was derived in the $f(\overline{\ourG})$ case~(\ref{fieldd4}). We can expand the right hand side, cancelling the second partial derivatives, to obtain
\begin{align}
 (\nabla_{\lambda} + T^{\sigma}{}_{\lambda \sigma}) \delta^{\lambda}_{[{\mu}} \partial_{ {\nu}]}  f_{,\overline{\ourB}}  &=  \frac{1}{2}\Big( T^{\lambda}{}_{ {  \nu \mu}} \partial_{\lambda}  f_{,\overline{\ourB}} + T^{\sigma}{}_{{ \mu} \sigma} \partial_{ { \nu}}  f_{,\overline{\ourB}} - T^{\sigma}{}_{{ \nu} \sigma} \partial_{{ \mu}}  f_{,\overline{\ourB}} \Big) \nonumber \\
  &= - \frac{1}{2}P^{\lambda}{}_{ { \nu \mu} } \partial_{\lambda}    f_{,\overline{\ourB}} \,,
  \label{ECmod5}
\end{align}
where the first torsion term comes from the skew covariant derivative terms. Putting this back into our skew-symmetric field equation~(\ref{fGB1}) we find
\begin{align}
  \Big(\overline{E}_{[\mu \nu]}{}^{\lambda} - P^{\lambda}{}_{{ \nu \mu}} \Big)
  \Big(\partial_{\lambda} f_{,\overline{\ourG}} - \partial_{\lambda} f_{,\overline{\ourB}}\Big) = 0 \,,
  \label{ECmod6}
\end{align}
where we again have $\overline{E}_{[\mu \nu]}{}^{\lambda} = P^{\lambda}{}_{ {\nu \mu} }$. If we use the object $\mathcal{E}_{\mu \nu}{}^{\lambda}$ introduced in~(\ref{fieldd5}), we can write the full field equation as 
\begin{multline}
  - \frac{1}{2} g_{\mu \nu} f +
  f_{,\overline{\ourG}} \big(\overline{G}_{\mu \nu} + \frac{1}{2} g_{\mu \nu} \overline{\ourG}\big) +
  \frac{1}{2} \mathcal{E}_{\mu \nu}{}^{\lambda} \partial_{\lambda} \big(f_{,\overline{\ourG}}-f_{,\overline{\ourB}} \big) +
  \frac{1}{2} g_{\mu \nu} f_{,\overline{\ourB}} \overline{\ourB}  \\ =
  \kappa \Big( {}^{(\overline{\Gamma})}T_{\mu \nu} + (\nabla_{\lambda} + T^{\sigma}{}_{\lambda \sigma}) \Delta^{\lambda}{}_{ { \nu \mu}} \Big) \,,
  \label{ECmod7}
\end{multline}
where the symmetric part contains the metric field equation and the skew part contains the connection equation.

The presence of the boundary terms in the form $f_{,\overline{\ourB}}$ in the second field equation make this model distinctly different to the previous modified Einstein-Cartan type theories. In Eqs.~(\ref{fieldd1c}) and~(\ref{fieldd2b}) the vanishing of the source term, that is $\Delta_{\mu[\nu \lambda]}=0$, implied the vanishing of the Palatini tensor which in turn implies the vanishing of the torsion tensor. One often says that torsion is non-dynamical in Einstein-Cartan theory.

Let us now set $\Delta_{\mu[\nu \lambda]}=0$ in Eq.~(\ref{ECmod2}). Using the expression of the Palatini tensor in terms of torsion, one can solve for the torsion tensor and finds the neat expression
\begin{align}
  T^{\mu}{}_{\lambda \nu} = \frac{1}{f_{,\overline{\ourG}}}
  \delta^{\mu}_{[\lambda} \partial_{\nu]} f_{,\overline{\ourB}} \,.
  \label{ECmod8}
\end{align}
Therefore, torsion does not vanish in general in source-free regions of spacetime. It should be recalled that the boundary term itself contains contributions from torsion which means that $\partial_{\mu} f_{,\overline{\ourB}}$ contains derivatives of the torsion tensor. Consequently, Eq.~(\ref{ECmod8}) is a partial differential equation for torsion. Using the chain rule we have
\begin{align}
  \partial_{\mu} f_{,\overline{\ourB}} = f_{,\overline{\ourB}\,\overline{\ourB}} \partial_{\mu}  \overline{\ourB} +
  f_{,\overline{\ourB}\,\overline{\ourG}} \partial_{\mu}  \overline{\ourG} \,,
  \label{ECmod9}
\end{align}
so that our previous expression~(\ref{ECmod8}) can also be written as
\begin{align}
  T^{\mu}{}_{\lambda \nu} = \frac{f_{,\overline{\ourB}\,\overline{\ourB}}}{f_{,\overline{\ourG}}}
  \delta^{\mu}_{[\lambda} \partial_{\nu]} \overline{\ourB} +
  \frac{f_{,\overline{\ourB}\,\overline{\ourG}}}{f_{,\overline{\ourG}}}
  \delta^{\mu}_{[\lambda} \partial_{\nu]} \overline{\ourG} \,.
  \label{ECmod10}
\end{align}
Neglecting the peculiar situations in which the partial derivative terms vanish, we can make the following general statements: $f(\overline{\ourG},\overline{\ourB})$ Einstein-Cartan type models have non-dynamical torsion if $f_{,\overline{\ourB}} = \mathrm{const.}$, which corresponds to functions linear in the boundary term. If $f_{,\overline{\ourB}} \neq \mathrm{const.}$ torsion becomes dynamical. 

As was already remarked at various occasions, these boundary terms play a crucial role in these modified theories of gravity. Models depending on non-linear functions of these boundary terms show distinctly different properties than models with a linear term, which is perhaps expected as an arbitrary function of a boundary term is no longer a boundary term.

\section{Cosmology}

\subsection{Brief introduction}

It is well known that the cosmological principle reduces the number of independent components of the metric tensor to just one free function, namely the scale factor $a(t)$. One often works with an arbitrary lapse function $N(t)$ which we generally set to one, $N(t)=1$. One can verify that this choice is consistent by noting that the conservation equation~(\ref{cons1}) is satisfied for this model for a general lapse function $N(t)$ in these coordinates. The spatially flat FLRW line element is given by
\begin{align}
  ds^2 = -dt^2 + a^2(t) (dx^2 + dy^2 +dz^2)\,.
  \label{cos1}
\end{align}
The torsion tensor is also strongly constrained when assuming isotropy and homogeneity~\cite{TSAMPARLIS197927}. The only allowed components of a cosmological torsion tensor are given by
\begin{align}
  T^1{}_{10} = T^2{}_{20} = T^3{}_{30} &=: h(t)/a^3(t) \,, \\
  T^1{}_{23} = T^2{}_{31} = T^3{}_{12} &=: k(t) \,.
  \label{cos2}
\end{align}
It will becomes clear shortly why the factor of $a^{-3}$ was included. Let us now consider our equation~(\ref{ECmod8}) assuming a cosmological setting where all objects are functions of time only. Then the right-hand side of~(\ref{ECmod8}) identically vanishes if all indices take spatial values $\{x,y,z\}$. Therefore, this source free equation immediately yields $T^1{}_{23} = T^2{}_{31} = T^3{}_{12} = 0$, so that we can set $k(t)=0$, and one is left with only a vector torsion contribution 
\begin{align}
  T_\mu = T^\sigma{}_{\sigma\mu} = \{3h/a^3,0,0,0\} \,.
  \label{cos2a}
\end{align}
Equation~(\ref{ECmod8}) simplifies into a single equation
\begin{align}
  \frac{h}{a^3} = T^{1}{}_{10} = T^2{}_{20} = T^3{}_{30} =
  \frac{1}{2}\frac{1}{f_{,\overline{\ourG}}}
  \partial_t f_{,\overline{\ourB}} \,.
  \label{cos3}
\end{align}
This cosmological spacetime only contains the torsion vector and this torsion vector has only one non-trivial component. Therefore the norm of this vector, which is a scalar, contains all the information about torsion of this spacetime
\begin{align}
  |T_{\rm vec}| = \sqrt{-g_{\mu\nu} T^\mu T^\nu} = \frac{3h}{a^3} \,.
\end{align}
In what follows it will be most natural to discuss the properties of $h/a^3$.

The boundary term $\overline{\ourB}$ with vanishing non-metricity contains a metric part $\ourB$ and a torsional contribution $B_T$, see Eq.~(\ref{B_scalars}). These are given by
\begin{align}
  \ourB &= \frac{1}{\sqrt{-g}}\partial_\nu
  \Bigl(\frac{1}{\sqrt{-g}} \partial_\mu(g g^{\mu\nu})\Bigr) = 18H^2 + 6\dot{H} \\
  B_T &= \frac{2}{\sqrt{-g}}\partial_\mu(\sqrt{-g} T^\mu) = -\frac{3}{a^3} \dot{h}\,.
  \label{cos4}
\end{align}
Here $T^\mu = T_\lambda{}^{\lambda\mu}$ and $H=\dot{a}/a$ is the Hubble function. It is the term $\sqrt{-g} T^\mu$ which motivated the above mentioned factor of $a^{-3}$ in~(\ref{cos2}). The torsion field equation now becomes
\begin{align}
  \frac{h}{a^3} = \frac{1}{2}\frac{1}{f_{,\overline{\ourG}}}
  \bigl(f_{,\overline{\ourB}\,\overline{\ourB}} \partial_t \overline{\ourB} +
  f_{,\overline{\ourB}\,\overline{\ourG}} \partial_t \overline{\ourG} \bigr) \,.
  \label{cos5}
\end{align}
It is already clear at this point that torsion can be non-trivial despite the absence of source terms for the torsion tensor. This is similar to the General Relativity setting where one typically finds vacuum solution with curvature.

\subsection{Quadratic boundary term model}

Let us propose to study specific models of the form
\begin{align}
  f(\overline{\ourG}, \overline{\ourB}) = \overline{\ourG} + F(\overline{\ourB}) \,,
\end{align}
with $F(\overline{\ourB})$ being an arbitrary function of the boundary term satisfying $F''(\overline{\ourB}) \neq 0$ such that torsion does not vanish. It turns out to be useful to make a redefinition of the torsion component $h(t) = \dot{a}a^2 + \mathfrak{h}(t)$. Then $h/a^3 = H + \mathfrak{h}/a^3$ which will simplify the subsequent field equations.

The complete cosmological field equations now take the following form
\begin{align}
  \frac{1}{2}F + 3\frac{\dot{\mathfrak{h}} }{a^3}F' + 3\frac{\mathfrak{h}^2}{a^6}  &= \rho \,,
  \label{cos5a} \\
  -\frac{1}{2} F - 3\frac{\dot{\mathfrak{h}}}{a^3} F' +
  \frac{12\mathfrak{h}}{a^6}\Bigl(\ddot{\mathfrak{h}}-3\frac{\dot{a}}{a}\dot{\mathfrak{h}}\Bigr) F'' + 2\frac{\dot{\mathfrak{h}}}{a^3} - \frac{\mathfrak{h}^2}{a^6} + 4\frac{\mathfrak{h}}{a^3}\frac{\dot{a}}{a} &= p = w \rho \,,
  \label{cos6}
\end{align}
where we assume the linear equation of state $p=w\rho$. The connection equation is 
\begin{align}
  \label{cos7}
  6\Bigl(\frac{\ddot{\mathfrak{h}}}{a^3} -
  3 \frac{\dot{a}}{a}\frac{\dot{\mathfrak{h}}}{a^3}\Bigr) F'' -
  2\frac{\mathfrak{h}}{a^3} 2- \frac{\dot{a}}{a} = 0 \,.
\end{align}
The dependence on $F(\overline{\ourB})$ and its derivatives can then be eliminated, leading to a \textit{Riccati}\footnote{A Riccati equation is a non-linear ODE of the form $y'(x) = q_0(x) + q_1(x) y + q_2(x) y^2$. In the above we have $q_1 = 0$.} differential equation relating torsion, the scale factor and the matter content
\begin{align}
  \label{Ricatti}
  \dot{\mathfrak{h}} = \frac{1}{2}(1+w) a^3 \rho - 3\frac{\mathfrak{h}^2}{a^3} \,.
\end{align}
It is noteworthy that this equation does not depend explicitly on the function $F(\overline{\ourB})$. This means torsion is uniquely determined once the scale factor and the matter evolution are known. However, these do depend on the specific model in question and hence will affect the form of torsion. 

At this point no further generic results can be extracted without specifying a concrete model. We choose one of the simplest possible settings, namely
\begin{align}
  F(\overline{\ourB}) = -\beta \overline{\ourB}^2 \,,
\end{align}
with $\beta > 0$. The chosen sign will become clear in what follows. The task at hand is to solve the system of equations~(\ref{cos5a})--(\ref{cos6}), together with the Riccati equation~(\ref{Ricatti}) and continuity equation for matter. Since the boundary term depends on torsion and the scale factor, one naturally has a system of two equations in two unknowns. Note that matter satisfies the standard conservation equation in our model so that $\rho = \rho_0/a^{3(1+w)}$. It turns out that one can eliminate torsion from the resulting equations and arrive at a single first order equation in the scale factor. For simplicity, we will only state this equation for $w=0$, but the general equation is of similar form, 
\begin{align}
  \frac{\dot{a}}{a} = \frac{\sqrt{2}}{2\sqrt{3\beta}} \frac{Y\sqrt{Y^2+2Y-3}}{Y^2+4Y-1}\,,
  \qquad Y^2 = 1+\frac{36\beta\rho_0}{a^3} \,.
  \label{ODE}
\end{align}
The introduction of the variable $Y$ simplifies the presentation of this key equation substantially. The main observation is that the right-hand side is a function of the scale factor only. Therefore, this equation is, in principle, separable. Perhaps surprisingly, the resulting integral can be expressed in terms of elementary functions. This equation cannot be solved for $a(t)$ explicitly but gives a closed form implicit solution. Some additional detail is given in Appendix~\ref{appODE}.

Despite the somewhat complicated form of~(\ref{ODE}) we can immediately make the following observations about solutions to this equation. First we make a series expansion assuming $a(t) \gg 1$, which corresponds to the late-time universe, and arrive at
\begin{align}
  \frac{\dot{a}}{a} = \frac{\sqrt{\rho_0}}{\sqrt{3}} \frac{1}{a^{3/2}} + O(a^{-5/2}) \,,
  \label{ODElate}
\end{align}
which is in agreement with a matter dominated universe. In particular, the late-time behaviour is independent of $\beta$ which strongly suggests that the boundary term affects the early-time dynamics of the universe only.

We now consider a series expansion of~(\ref{ODE}) assuming $a(t) \ll 1$, which gives
\begin{align}
  \frac{\dot{a}}{a} = \frac{\sqrt{2}}{3\sqrt{3}\sqrt{\beta}} + O(a^{3/2}) \,,
  \label{ODEearly}
\end{align}
giving a constant Hubble function at early times. Writing $a = a_0 \exp(\lambda t)$ for the early universe, one finds the very neat result
\begin{align}
  \lambda = \frac{\sqrt{2}}{3\sqrt{3}\sqrt{\beta}} \,.
  \label{ODEearly2}
\end{align}
Our model can thus lead to a large amount of inflation for small values of $\beta$. Let us also note that the previous equation can easily be found for general $w$, which gives the expression
\begin{align}
  \lambda = \frac{\sqrt{2}}{\sqrt{3}\sqrt{\beta}}\frac{1}{3\sqrt{1-w^2}} \,.
  \label{ODEearly3}
\end{align}
Whilst the value of $\lambda$ is affected by the matter equation of state $w$, we can safely state that reasonable matter choices have no qualitative impact on this inflationary epoch.

This early-time inflation will nonetheless yield a late-time matter dominated universe whose expansion is independent of $\beta$. It is reasonable to expect that the introduction of the standard cosmological constant into this model will yield late-time accelerated expansion. Self-accelerating solutions with propagating torsion have been found in other theories with higher powers of curvature scalars even without spin/hypermomentum sources, e.g., in the context of massive gravity~\cite{Nikiforova:2016ngy,Nikiforova:2018pdk,Deffayet:2011uk}. However, given the simplicity of the model studied here, this is a most surprising result.

\begin{figure}[!htb] 
\centering
\includegraphics[width=.75\linewidth]{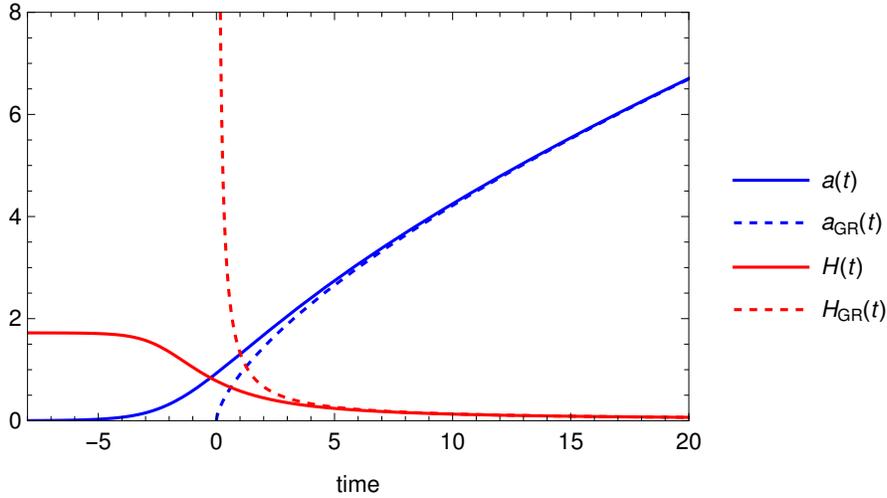}  
\caption{Solution of the field equations showing the scale factor $a(t)$ and Hubble factor $H(t)$ for a matter dominated universe. The dashed lines show the standard GR solutions. We note that the Hubble function was multiplied by a factor of 10 to improve the plot. The following numerical values were used: $w=0$, $\rho_0 =1$, $\kappa=1$, $\beta=1/10$. Other values give qualitatively similar results.}
\label{fig1a}
\end{figure}

Figure~\ref{fig1a} shows the evolution of the scale factor and the Hubble parameter for a matter dominated universe, in agreement with the above asymptotic discussions. We also show the relevant results of standard cosmology, as dashed lines. Contrary to standard cosmology where the Hubble function diverges as $t \to 0$ we find that $H \to \lambda$ as $t \to -\infty$. This discussion can also be repeated for a radiation equation of state where the epoch of early-time inflation would then be followed by a radiation dominated epoch.

\begin{figure}[!hbt] 
\centering
\includegraphics[width=.80\linewidth]{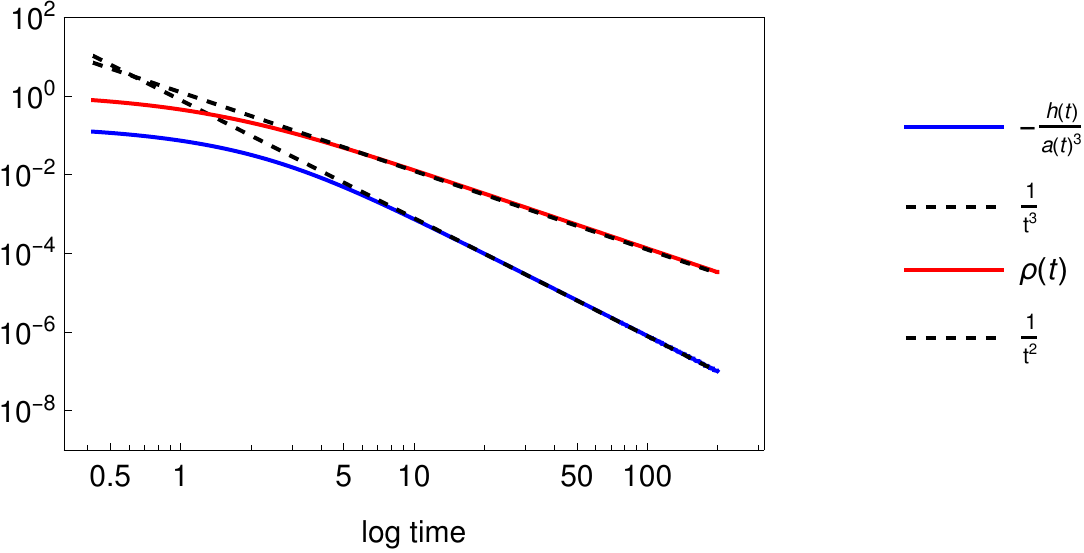}  
\caption{Log-log plot showing torsion $-h(t)/a^3$ (blue) and the energy density $\rho(t)$ (red) for a matter dominated universe together with the $1/t^3$ and $1/t^2$, shown as dashed lines. The following numerical values were used: $w=0$, $\rho_0 =1$, $\kappa=1$, $\beta=1/10$. Other values give qualitatively similar results.}
\label{fig2a}
\end{figure}

Next, let us turn to the behaviour of torsion. Since the late-time behaviour of the scale factor and matter agree with the GR results, we can assume $\rho \propto a^{-3} \propto t^{-2}$ and $a \propto t^{2/3}$ for late times, again assuming $w=0$. Using this input in~(\ref{Ricatti}) we can solve the Ricatti equation, explicitly or numerically. For the matter dominated epoch we find $-h/a^3 \propto 1/t^3 \propto \rho^{3/2}$ which, using the form of $a(t)$, also yields $-h \propto \sqrt{\rho}$. This is confirmed in Figure~\ref{fig2a} where we display a log-log plot to emphasise the scaling behaviour of the late-time solution.

\begin{figure}[!htb] 
\centering
\includegraphics[width=.80\linewidth]{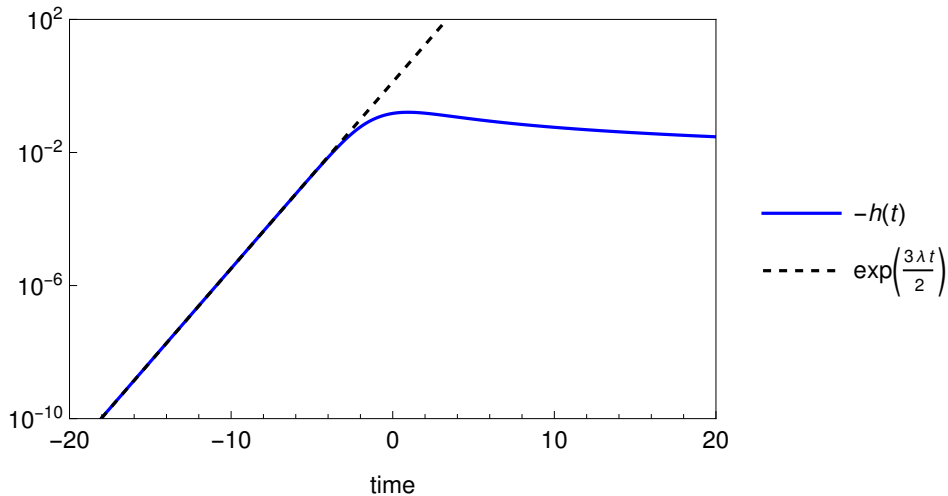}  
\caption{Log plot showing torsion $-h(t)$ and the solution $\exp(3\lambda t/2)$, again for a matter dominated universe. The following numerical values were used: $w=0$, $\rho_0 =1$, $\kappa=1$, $\beta=1/10$. Other values give qualitatively similar results.}
\label{fig2b}
\end{figure}

Likewise, we can also study the Ricatti equation assuming $a \propto \exp(\lambda t)$. As before one has $\rho \propto a^{-3}$ because the equation of state is unchanged. This gives $-h(t) \propto \exp(3\lambda t/2)$, which we can again verify using the numerical solutions, shown in Figure~\ref{fig2b}. Rather interestingly, torsion again scales with the scale factor, and perhaps even more noteworthy, it thus scales with the matter. In particular one finds $-h(t) \propto 1/\sqrt{\rho}$ which is the inverse relation when compared to the late-time solution.

\section{Conclusions and discussions}
\label{sec:disc}

The primary objective of this work was to generalise our previous results in the metric-affine or Palatini way. By this we mean gravity models where the metric and the connection are treated as \emph{a priori} independent variables. Within this framework one can recover theories like Einstein-Cartan theory or General Relativity via the introduction of suitable Lagrange multipliers. From our point of view this is the most elegant approach to consider particular models without torsion or without non-metricity. Teleparallel models can also be studied by setting curvature to zero, again via a Lagrange multiplier. While other approaches can work, directly setting some quantities to zero in the action can prove problematic. This can be seen succinctly in Section~\ref{subsec:metricaffineGR}. One might be tempted to set non-metricity and the Lagrange multiplier to zero, however, this is inconsistent because the Lagrange multiplier is in fact non-vanishing, as already commented on in that section.

The theory proposed here, based on the non-covariant decomposition of the affine Ricci scalar, breaks diffeomorphism invariance and therefore allows for a number of interesting new gravitational models to be studied. To maintain the consistency of the variational methods procedure, we enforced that the total action be invariant under arbitrary coordinate transformations. This introduces a set of `constraint equations' or conservation equations that must be satisfied for all solutions, which can alternatively be viewed as picking out appropriate coordinates. Here, as well as the examples studied in our previous work~\cite{Boehmer:2021aji}, the gravitational sector satisfied these equations, implying the usual conservation laws for the matter energy-momentum tensor.

After setting up the entire framework so that metric-affine theories can be studied, we were particularly interested in models in an Einstein-Cartan geometry with propagating torsion. We demonstrated that torsion can exists in certain settings, even in the absence of sources. This was achieved by considering actions containing non-linear boundary terms, at which point they can no longer be seen as boundary terms. The explicit conditions needed to produce propagating torsion were derived, and the presence of these modified boundary terms were found to be essential.

Using a simple quadratic term, we considered cosmological solutions assuming the standard FLRW framework. In this setting the metric only contains a single unknown function while the torsion tensor is reduced to two independent components, a vector torsion piece and an axial torsion piece. In our specific model the axial torsion component had to vanish identically to satisfy the cosmological field equations, leaving us with only one additional function when compared to General Relativity. The resulting field equations are relatively easy to deal with and it is straightforward to study their asymptotic behaviour for early and late times. We found late-time solutions in agreement with standard cosmology, whilst the early-time solutions were affected by our modified non-linear terms. In particular, we found an early-time epoch of inflation, driven solely by our geometric torsion component. Our solutions offers a smooth transition from an epoch of accelerated expansion to an epoch of matter (or radiation) domination. This is achieved using a relatively simple modified theory of gravity based primarily on the Einstein action, without the need of additional fields.

It should again be emphasised that our model generally breaks diffeomorphism invariance, with a new set of constraint equations resulting from requiring the action to be invariant under infinitesimal coordinate transformations. We interpreted these equations as constraints on the allowed coordinates of our metric and connection equations. In the cosmological model studied above, using Cartesian coordinates and the standard FLRW symmetries, these constraint equations are automatically satisfied. Consequently, the energy-momentum tensor is covariantly conserved, despite the lack of invariance of the action.

The cosmological solutions were studied in the Einstein-Cartan framework with vanishing non-metricity, but the full metric-affine theory could easily be incorporated into the study. Similarly, our general field equations and conservation equations allow for the possibility of connection source terms, the hypermomentum current. This may lead to a more interesting, albeit more complicated, set of solutions. One could also generalise the approach to allow for non-minimal couplings between the geometrical pseudo-scalars and the matter content, at which point the usual energy-momentum conservation laws would be modified. While non-minimal couplings between matter and geometry are straightforward to formulate at the level of the action, it is difficult to make qualitative statements about the effects of such terms without performing a detailed investigation. Moreover, we lack the knowledge of any guiding principles to select certain couplings over others, other than perhaps a preference for simpler models. Having said this, the minimally coupled theory presented here can be seen as the most obvious modification based on the natural decomposition of the Ricci scalar into its bulk and boundary parts within the metric-affine framework.

\subsection*{Acknowledgements}

We thank Yuri Obukhov for valuable discussions. Erik Jensko is supported by EPSRC Doctoral Training Programme (EP/R513143/1).

\appendix

\section{Explicit variational calculations}

\subsection{Derivation of Einstein field equations}
\label{A1}

In this Appendix we show the explicit derivation of the metric-affine field equations from the Einstein-Palatini action~(\ref{Einstein_var}) and the field equations from the modified action~(\ref{full_variation}). We first briefly show that the bulk term $\overline{\ourG}$ can be written in an equivalent form with the Palatini tensor~(\ref{Palatini}) by using the relations  
 $\partial_{\lambda} g^{\mu \nu} = Q_{\lambda}{}^{\mu \nu} - 2 \overline{\Gamma}{}^{(\mu}_{\lambda \eta} g^{\nu) \eta}$ and $\overline{\Gamma}{}^{\mu}_{\nu \lambda} = T^{\mu}{}_{\nu \lambda} +  \overline{\Gamma}{}^{\mu}_{\lambda \nu} $ to obtain
\begin{align}
\overline{\ourG} &:= g^{\mu \lambda} \big( \overline{\Gamma}{}^{\kappa}_{\kappa \rho} \overline{\Gamma}{}^{\rho}_{\mu \lambda} - \overline{\Gamma}{}^{\kappa}_{\mu \rho} \overline{\Gamma}{}^{\rho}_{\kappa \lambda} \big) + \big(\overline{\Gamma}{}^{\mu}_{\mu \lambda} \delta^{\nu}_{\kappa} - \overline{\Gamma}{}^{\nu}_{\kappa \lambda} \big) \big( \partial_{\nu} g^{\kappa \lambda} - \frac{1}{2} g_{\alpha \beta} g^{\kappa \lambda} \partial_{\nu} g^{\alpha \beta} \big)  \nonumber \\
&= g^{\mu \lambda} \big( \overline{\Gamma}{}^{\kappa}_{\kappa \rho} \overline{\Gamma}{}^{\rho}_{\mu \lambda} - \overline{\Gamma}{}^{\kappa}_{\mu \rho} \overline{\Gamma}{}^{\rho}_{\kappa \lambda} \big) + 2 g^{\mu \lambda} \big( \overline{\Gamma}{}^{\kappa}_{\rho \lambda} \overline{\Gamma}{}^{\rho}_{\mu \kappa} - \overline{\Gamma}{}^{\kappa}_{\mu \lambda} \overline{\Gamma}{}^{\rho}_{\rho \kappa}  \big)  \nonumber \\
&+ \overline{\Gamma}{}^{\kappa}_{\mu \lambda} \big(-Q_{\kappa}{}^{\mu \lambda} + \delta^{\mu}_{\kappa} Q_{\rho}{}^{\rho \lambda} + \frac{1}{2} g^{\mu \lambda} Q_{\kappa \rho}{}^{\rho} - \frac{1}{2} \delta^{\mu}_{\kappa} Q^{\lambda \rho}{}_{\rho} + T^{\mu}{}_{\kappa}{}^{\lambda} - g^{\mu \lambda} T^{\rho}{}_{\kappa \rho} + \delta^{\mu}_{\kappa} T^{\rho \lambda}{}_{\rho} \big) \nonumber \nonumber \\
&= g^{\mu \lambda} \big( \overline{\Gamma}{}^{\kappa}_{\rho \lambda} \overline{\Gamma}{}^{\rho}_{\mu \kappa} - \overline{\Gamma}{}^{\kappa}_{\mu \lambda} \overline{\Gamma}{}^{\rho}_{\rho \kappa}  \big) + \overline{\Gamma}{}^{\kappa}_{\mu \lambda}  P^{\mu \lambda}{}_{\kappa} \,.
\label{Gbar-nice1}
\end{align}
Note that the quadratic terms have flipped sign in going from the first to the final line. This form is given in equation~(\ref{G3}). 

Looking at the special case of vanishing non-metricity, we can swap the indices of the connection terms to introduce torsion tensors that cancel terms in the Palatini tensor
\begin{align}
  \overline{\ourG} &= g^{\mu \lambda} \big( \overline{\Gamma}{}^{\kappa}_{\rho \lambda} \overline{\Gamma}{}^{\rho}_{\mu \kappa} - \overline{\Gamma}{}^{\kappa}_{\mu \lambda} \overline{\Gamma}{}^{\rho}_{\rho \kappa}  \big) + \overline{\Gamma}{}^{\kappa}_{\mu \lambda}  P^{\mu \lambda}{}_{\kappa} \nonumber \\
& \stackrel{Q=0}{=} g^{\mu \lambda} \big( \overline{\Gamma}{}^{\kappa}_{\rho \lambda} \overline{\Gamma}{}^{\rho}_{\kappa \mu} - \overline{\Gamma}{}^{\kappa}_{\mu \lambda} \overline{\Gamma}{}^{\rho}_{\kappa \rho}  \big) - \overline{\Gamma}{}^{\lambda}_{\lambda \mu} T^{\rho}{}_{\rho}{}^{\mu} \,.
\end{align}
This is the form of the bulk term in an Einstein-Cartan space, for example see~\cite[see Eq.~(3.20)]{FH1976}.

Returning to the variations, the metric variation of the Einstein-Palatini action is straightforward but tedious. A direct calculation gives
\begin{align}
  \delta_g S &= \frac{1}{2\kappa} \int \delta (\sqrt{-g} \overline{\ourG}) \, d^4 x
  \nonumber \\
  &\begin{multlined}
     =\frac{1}{2\kappa} \int \biggl\{- \frac{1}{2} \delta g^{\mu \nu} g_{\mu \nu} \overline{\ourG} +
     \Big[ \delta g^{\mu \nu}  \big( \overline{\Gamma}^{\kappa}_{\kappa \rho} \overline{\Gamma}^{\rho}_{\mu \nu} - \overline{\Gamma}^{\kappa}_{\mu \rho} \overline{\Gamma}^{\rho}_{\kappa \nu} \big)  \\
       + \big(\overline{\Gamma}^{\mu}_{\mu \lambda} \delta^{\nu}_{\kappa} - \overline{\Gamma}^{\nu}_{\kappa \lambda} \big)\big(\partial_{\nu} \delta g^{\kappa \lambda} - \frac{1}{2} \delta (g_{\alpha \beta} g^{\kappa \lambda} \partial_{\nu} g^{\alpha \beta})\big) \Big]\biggr\} \sqrt{-g} d^4x
   \end{multlined}
  \nonumber \\
  &\begin{multlined}[b]
     =\frac{1}{2\kappa}  \int \sqrt{-g} \delta g^{\mu \nu} \Big( - \frac{1}{2} g_{\mu \nu} \overline{\ourG} + \big( \overline{\Gamma}^{\kappa}_{\kappa \rho} \overline{\Gamma}^{\rho}_{\mu \nu} - \overline{\Gamma}^{\kappa}_{\mu \rho} \overline{\Gamma}^{\rho}_{\kappa \nu} \big)  \Big) d^4x   \\
     + \mathrm{boundary\ terms} - \frac{1}{2\kappa} \int \delta g^{\kappa \lambda} \partial_{\nu} \Bigl(\sqrt{-g} \big(\overline{\Gamma}^{\mu}_{\mu \lambda} \delta^{\nu}_{\kappa} - \overline{\Gamma}^{\nu}_{\kappa \lambda} \big)\Bigr) d^4x
     \\
     - \frac{1}{2\kappa}\int \frac{1}{2}  \sqrt{-g} \big(\overline{\Gamma}^{\mu}_{\mu \lambda} \delta^{\nu}_{\kappa} - \overline{\Gamma}^{\nu}_{\kappa \lambda} \big) 
     \big( \delta g_{\alpha \beta} g^{\kappa \lambda} \partial_{\nu} g^{\alpha \beta} + g_{\alpha \beta} \delta g^{\kappa \lambda} \partial_{\nu} g^{\alpha \beta} \big) d^4x
     \\
     - \mathrm{boundary\ terms} + \frac{1}{2\kappa} \int \frac{1}{2} \delta g^{\alpha \beta} \partial_{\nu}\Bigl(\sqrt{-g} \big(\overline{\Gamma}^{\mu}_{\mu \lambda} \delta^{\nu}_{\kappa} - \overline{\Gamma}^{\nu}_{\kappa \lambda} \big)  g_{\alpha \beta} g^{\kappa \lambda} \Bigr) d^4 x \,,
   \end{multlined} 
\end{align}
where the boundary terms come from integration by parts. Discarding these boundary terms leaves
\begin{multline}
= \frac{1}{2\kappa} \int \sqrt{-g} \delta g^{\sigma \gamma} \Big[ -\frac{1}{2} g_{\sigma \gamma} g^{\mu \lambda} \big( \overline{\Gamma}^{\kappa}_{\kappa \rho} \overline{\Gamma}^{\rho}_{\mu \lambda} - \overline{\Gamma}^{\kappa}_{\mu \rho} \overline{\Gamma}^{\rho}_{\kappa \lambda} \big)
-\frac{1}{2} g_{\sigma \gamma} \big(\overline{\Gamma}^{\mu}_{\mu \lambda} \delta^{\nu}_{\kappa} - \overline{\Gamma}^{\nu}_{\kappa \lambda} \big) \big( \partial_{\nu} g^{\kappa \lambda} - \frac{1}{2} g_{\alpha \beta} g^{\kappa \lambda} \partial_{\nu} g^{\alpha \beta} \big)  \\
+  \big( \overline{\Gamma}^{\kappa}_{\kappa \rho} \overline{\Gamma}^{\rho}_{\sigma \gamma} - \overline{\Gamma}^{\kappa}_{\sigma \rho} \overline{\Gamma}^{\rho}_{\kappa \gamma} \big) - \frac{1}{\sqrt{-g}} \partial_{\nu} (\sqrt{-g} \big(\overline{\Gamma}^{\mu}_{\mu \gamma} \delta^{\nu}_{\sigma} - \overline{\Gamma}^{\nu}_{\sigma \gamma} \big)) 
- \frac{1}{2} \big(\overline{\Gamma}^{\mu}_{\mu \lambda} \delta^{\nu}_{\kappa} - \overline{\Gamma}^{\nu}_{\kappa \lambda} \big) g^{\kappa \lambda} \partial_{\nu} g_{\sigma \gamma}   \\
-\frac{1}{2} (\overline{\Gamma}^{\mu}_{\mu \gamma} \delta^{\nu}_{\sigma} - \overline{\Gamma}^{\nu}_{\sigma \gamma} \big) g_{\alpha \beta} \partial_{\nu} g^{\alpha \beta}
+ \frac{1}{2} \frac{1}{\sqrt{-g}} \partial_{\nu}\Big(\sqrt{-g} \big(\overline{\Gamma}^{\mu}_{\mu \lambda} \delta^{\nu}_{\kappa} - \overline{\Gamma}^{\nu}_{\kappa \lambda} \big)  g_{\sigma \gamma} g^{\kappa \lambda} \Big) 
 \Big] \, d^4 x \,,
\end{multline}
where we have used relations such as $\delta g_{\alpha \beta} \partial_{\nu} g^{\alpha \beta} = - \delta g^{\sigma \gamma} g_{\sigma \alpha} g_{\gamma \beta}  \partial_{\nu} g^{\alpha \beta} = \delta g^{\sigma \gamma} \partial_{\nu} g_{\sigma \gamma}$. Introducing the shorthand notation $\tilde{\Gamma}^{\nu}_{\kappa \lambda} := \overline{\Gamma}^{\mu}_{\mu \lambda} \delta^{\nu}_{\kappa} - \overline{\Gamma}^{{\nu}}_{\kappa \lambda}$ and expanding the partial derivatives, we have 
\begin{multline}
  = \frac{1}{2\kappa} \int \sqrt{-g} \delta g^{\sigma \gamma} \bigg[ -\frac{1}{2} g_{\sigma \gamma} g^{\mu \lambda} \big( \overline{\Gamma}^{\kappa}_{\kappa \rho} \overline{\Gamma}^{\rho}_{\mu \lambda} - \overline{\Gamma}^{\kappa}_{\mu \rho} \overline{\Gamma}^{\rho}_{\kappa \lambda} \big) + \big( \overline{\Gamma}^{\kappa}_{\kappa \rho} \overline{\Gamma}^{\rho}_{\sigma \gamma} - \overline{\Gamma}^{\kappa}_{\sigma \rho} \overline{\Gamma}^{\rho}_{\kappa \gamma} \big) \\
    - \frac{1}{2} g_{\sigma \gamma} \tilde{\Gamma}^{\nu}_{\kappa \lambda} \partial_{\nu} g^{\kappa \lambda} + \frac{1}{4} g_{\sigma \gamma} \tilde{\Gamma}^{\nu}_{\kappa \lambda} g_{\alpha \beta} g^{\kappa \lambda} \partial_{\nu} g^{\alpha \beta} + \frac{1}{2} \tilde{\Gamma}^{\nu}_{\sigma \gamma} g_{\alpha \beta} \partial_{\nu} g^{\alpha \beta} \\ - \partial_{\nu} \tilde{\Gamma}^{\nu}_{\sigma \gamma} - \frac{1}{2}\tilde{\Gamma}^{\nu}_{\kappa \lambda} g^{\kappa \lambda} \partial_{\nu} g_{\sigma \gamma} - \frac{1}{2} \tilde{\Gamma}^{\nu}_{\sigma \gamma} g_{\alpha \beta} \partial_{\nu} g^{\alpha \beta} - \frac{1}{4} g_{\sigma \gamma} \tilde{\Gamma}^{\nu}_{\kappa \lambda} g_{\alpha \beta} g^{\kappa \lambda} \partial_{\nu} g^{\alpha \beta} \\ + \frac{1}{2} \partial_{\nu} \tilde{\Gamma}^{\nu}_{\kappa \lambda} g_{\sigma \gamma} g^{\kappa \lambda} + \frac{1}{2}\tilde{\Gamma}^{\nu}_{\kappa \lambda} g^{\kappa \lambda} \partial_{\nu} g_{\sigma \gamma} + \frac{1}{2} \tilde{\Gamma}^{\nu}_{\kappa \lambda} g_{\sigma \gamma} \partial_{\nu} g^{\kappa \lambda} \bigg]  d^4x \,,
\end{multline}
which simplifies to
\begin{align}
&\begin{multlined}
= \frac{1}{2\kappa}  \int \sqrt{-g} \delta g^{\sigma \gamma} \Big[ \big( \overline{\Gamma}^{\kappa}_{\kappa \rho} \overline{\Gamma}^{\rho}_{\sigma \gamma} - \overline{\Gamma}^{\kappa}_{\sigma \rho} \overline{\Gamma}^{\rho}_{\kappa \gamma} \big)  -\frac{1}{2} g_{\sigma \gamma} g^{\mu \lambda} \big( \overline{\Gamma}^{\kappa}_{\kappa \rho} \overline{\Gamma}^{\rho}_{\mu \lambda} - \overline{\Gamma}^{\kappa}_{\mu \rho} \overline{\Gamma}^{\rho}_{\kappa \lambda} \big) \nonumber  \\
+ \partial_{\nu} \overline{\Gamma}^{\nu}_{\sigma \gamma} - \partial_{\sigma} \overline{\Gamma}^{\lambda}_{\lambda \gamma} -\frac{1}{2} g_{\sigma \gamma} g^{\kappa \lambda}(\partial_{\nu} \overline{\Gamma}^{\nu}_{\kappa \lambda} - \partial_{\kappa} \overline{\Gamma}^{\mu}_{\mu \lambda}) \Big] d^4x 
 \end{multlined}
  \\&
  = \frac{1}{2\kappa} \int \sqrt{-g} \delta g^{\sigma \gamma} \overline{G}_{\sigma \gamma} \, d^4x \,.
\end{align}
Hence we have obtained the metric-affine Einstein tensor from the metric variation of the bulk term.

The connection variations simply give rise to the Palatini tensor~(\ref{Palatini}), which we show here for completeness. Splitting the calculation into two parts, the variation of the quadratic part of $\overline{\ourG}$ and the part containing derivatives of the metric
\begin{align}
\frac{\delta \overline{\ourG}_{\rm{quad}}}{\delta \overline \Gamma} &= \delta \overline{\Gamma}{}^{\gamma}_{\alpha \beta} \Big( \delta^{\alpha}_{\gamma} g^{\mu \lambda} \overline{\Gamma}{}^{\beta}_{\mu \lambda} 
+ g^{\alpha \beta} \overline{\Gamma}{}^{\kappa}_{\kappa \gamma} - g^{\alpha \kappa}\overline{\Gamma}{}^{\beta}_{\gamma \kappa} - g^{\kappa \beta}\overline{\Gamma}{}^{\alpha}_{\kappa \gamma}
\Big ) \,,  \\
\frac{\delta \overline{\ourG}_{\rm{other}} }{\delta \overline \Gamma} &=  \delta \overline{\Gamma}{}^{\gamma}_{\alpha \beta} \Big(\delta^{\alpha}_{\gamma} \partial_{\mu}g^{\mu \beta} - \partial_{\gamma} g^{\alpha \beta} +\frac{1}{2} g_{\mu \nu} g^{\alpha \beta} \partial_{\gamma}g^{\mu \nu} - \frac{1}{2} \delta^{\alpha}_{\gamma} g_{\mu \nu} g^{\kappa \beta} \partial_{\kappa} g^{\mu \nu} \Big ) \,.
\end{align}
Rewriting the partial derivatives of the metric in terms of non-metricity and connection terms, along with some cancellations, allows us to write the total variation as 
\begin{align}
\frac{\delta \overline{\ourG}}{\delta \overline \Gamma} &=  \frac{\delta \overline{\ourG}_{\rm{quad}}}{\delta \overline \Gamma} +\frac{\delta \overline{\ourG}_{\rm{other}} }{\delta \overline \Gamma} \nonumber  \\
&\begin{multlined}= \delta \overline{\Gamma}{}^{\gamma}_{\alpha \beta} \Big[ \delta^{\alpha}_{\gamma} Q_{\lambda}{}^{\lambda \beta} - Q_{\gamma}{}^{\alpha \beta} + \frac{1}{2}g^{\alpha \beta} Q_{\gamma}{}^{\lambda}{}_{\lambda} - \frac{1}{2} \delta^{\alpha}_{\gamma} Q^{\beta \lambda}{}_{\lambda}  \\
+ \delta^{\alpha}_{\gamma} g^{\mu \beta} \big( \overline{\Gamma}{}^{\lambda}_{\mu \lambda} -  \overline{\Gamma}{}^{\lambda}_{\lambda \mu} \big) + g^{\alpha \beta}\big( \overline{\Gamma}{}^{\lambda}_{\lambda \gamma} - \overline{\Gamma}{}^{\lambda}_{\gamma \lambda}\big) + g^{\mu \beta} \big( \overline{\Gamma}{}^{\alpha}_{\gamma \mu} -  \overline{\Gamma}{}^{\alpha}_{\mu \gamma} \big)
\Big]
\end{multlined} \nonumber \\
&= \delta \overline{\Gamma}{}^{\gamma}_{\alpha \beta} \Big[ \delta^{\alpha}_{\gamma} Q_{\lambda}{}^{\lambda \beta} - Q_{\gamma}{}^{\alpha \beta} + \frac{1}{2}g^{\alpha \beta} Q_{\gamma}{}^{\lambda}{}_{\lambda} - \frac{1}{2} \delta^{\alpha}_{\gamma} Q^{\beta \lambda}{}_{\lambda} + \delta^{\alpha}_{\gamma} T^{\lambda \beta}{}_{\lambda} + g^{\alpha \beta} T^{\lambda}{}_{\lambda \gamma} + T^{\alpha}{}_{\gamma}{}^{\beta}  \Big] \,,
\end{align}
where in the final line we used the definition of the torsion tensor.
We recognise this object in the brackets to be the Palatini tensor~(\ref{Palatini}).
Substituting this into the action gives
\begin{align}
\delta_{\overline{\Gamma}}S =  \frac{1}{2\kappa}  \int \sqrt{-g} \delta_{\overline{\Gamma}} (\overline{\ourG}) \, d^4x  = \frac{1}{2\kappa}  \int \sqrt{-g} \delta \overline{\Gamma}{}^{\lambda}_{\mu \nu} P^{\mu \nu}{}_{\lambda} d^4 x \,.
\end{align}

\subsection{Derivation of modified field equations}
\label{app-a2}

Now let us derive the field equations for the modified action
\begin{align} \label{Appendix_var}
\delta S &= \frac{1}{2\kappa }  \int \Big[ \delta \sqrt{-g} f(\overline{\ourG}, \overline{\ourB}) + \sqrt{-g} \delta f(\overline{\ourG}, \overline{\ourB}) \Big] d^4x \nonumber \\
&= \frac{1}{2\kappa}  \int \sqrt{-g} \Big[  -\frac{1}{2} g_{\mu \nu} \delta g^{\mu \nu} f(\overline{\ourG}, \overline{\ourB}) + \frac{\partial f(\overline{\ourG}, \overline{\ourB})}{\partial \overline{\ourG}}  \delta  \overline{\ourG} + \frac{\partial f(\overline{\ourG}, \overline{\ourB})}{\partial \overline{\ourB}}  \delta  \overline{\ourB} \Big] d^4x \,.
\end{align}
Using the previous calculations, and again using the shorthand notation $\tilde{\Gamma}^{\nu}_{\kappa \lambda} := \overline{\Gamma}^{\mu}_{\mu \lambda} \delta^{\nu}_{\kappa} - \overline{\Gamma}^{{\nu}}_{\kappa \lambda}$, the variation of the bulk term is
\begin{multline}
\delta \overline{\ourG} = \delta g^{\mu \lambda} \big( \overline{\Gamma}^{\kappa}_{\kappa \rho} \overline{\Gamma}^{\rho}_{\mu \lambda} - \overline{\Gamma}^{\kappa}_{\mu \rho} \overline{\Gamma}^{\rho}_{\kappa \lambda} \big) + \tilde{\Gamma}^{\nu}_{\kappa \lambda} \partial_{\nu} \delta g^{\lambda \kappa} 
 - \frac{1}{2} \tilde{\Gamma}^{\nu}_{\lambda \kappa} g_{\alpha \beta} g^{\kappa \lambda} \partial_{\nu} \delta g^{\alpha \beta} \\
 - \frac{1}{2} \delta g_{\alpha \beta} g^{\kappa \lambda} \partial_{\nu} g^{\alpha \beta} \tilde{\Gamma}^{\nu}_{\kappa \lambda} - \frac{1}{2} \delta g^{\kappa \lambda} g_{\alpha \beta} \partial_{\nu} g^{\alpha \beta} \tilde{\Gamma}^{\nu}_{\kappa \lambda} + \delta \overline{\Gamma}{}^{\lambda}_{\mu \nu} P^{\mu \nu}{}_{\lambda} \,,
\end{multline}
where the final two terms on the first line will require integration by parts. We can write this in a nicer form by recalling that for $f(\overline{\ourG}) = \overline{\ourG}$ the metric variations, along with the variation of the metric determinant, give the metric-affine Einstein tensor. The only additional terms for the modified action will come from integration by parts when the partial derivatives hits the $f_{, \overline{\ourG}}$ term in~(\ref{Appendix_var}). In other words, we have
\begin{multline}
  \label{A_G_final}
  \int \sqrt{-g}  f_{, \overline{\ourG}} \delta_g \overline{\ourG} \, d^4 x =  \int \sqrt{-g} \Big( \delta g^{\mu \nu} f_{, \overline{\ourG}}  \overline{G}_{\mu \nu} + \frac{1}{2} \delta g^{\mu \nu} f_{, \overline{\ourG}}  g_{\mu \nu} \overline{\ourG} \\+ \delta g^{\mu \nu} \partial_{\lambda} f_{, \overline{\ourG}} \big(
  \frac{1}{2} \tilde{\Gamma}^{\lambda}_{\rho \kappa} g_{\mu \nu} g^{\kappa \rho}
  -  \tilde{\Gamma}^{\lambda}_{\mu \nu}
   \big)
  \Big) d^4 x  \,.
\end{multline}
Also recall that the bulk term can be written in the form given in~(\ref{G2}) where all the derivative terms are included in $\frac{1}{2} \partial_{\lambda} g^{\mu \nu}  \overline{E}_{\mu \nu}{}^{\lambda}$. Therefore it must be the case that this final term is proportional to $ \overline{E}_{\mu \nu}{}^{\lambda}$, which is easily verified
 \begin{align} \label{E_verify}
\delta g^{\mu \nu} \partial_{\lambda} f_{, \overline{\ourG}} \Big(
  \frac{1}{2} \tilde{\Gamma}^{\lambda}_{\rho \kappa} g_{\mu \nu} g^{\kappa \rho}
  -  \tilde{\Gamma}^{\lambda}_{\mu \nu} \Big) &= \delta g^{\mu \nu} \partial_{\lambda} f_{, \overline{\ourG}} \Big( \frac{1}{2} g_{\mu \nu} g^{\kappa \lambda} \overline{\Gamma}{}^{\rho}_{\rho \kappa} - \frac{1}{2} g_{\mu \nu} g^{\kappa \lambda}  \overline{\Gamma}{}^{\lambda}_{\rho \kappa} +  \overline{\Gamma}{}^{\lambda}_{\mu  \nu} - \delta^{\lambda}_{\mu}  \overline{\Gamma}{}^{\rho}_{\rho \nu} \Big) \nonumber \\
  &= \frac{1}{2} \delta g^{\mu \nu} \partial_{\lambda} f_{, \overline{\ourG}}  \overline{E}_{\mu \nu}{}^{\lambda} \,.
\end{align}
In total, for the bulk term we have
 \begin{align}
\int \sqrt{-g}  f_{, \overline{\ourG}} \delta \overline{\ourG} \, d^4 x =  \int \sqrt{-g}  \delta g^{\mu \nu}  \Big(f_{, \overline{\ourG}}  \overline{G}_{\mu \nu} + \frac{1}{2} f_{, \overline{\ourG}}  g_{\mu \nu} \overline{\ourG} +  \frac{1}{2} \partial_{\lambda} f_{, \overline{\ourG}} \overline{E}_{\mu \nu}{}^{\lambda}   \Big) d^4 x  \,.
\end{align}

For the boundary term~(\ref{B}) we have
\begin{multline}
\delta  \overline{\ourB} = \frac{1}{2} g_{\alpha \beta} \delta g^{\alpha \beta}  \overline{\ourB}  + \frac{1}{\sqrt{-g}} \partial_{\kappa} \Bigg[\sqrt{-g}  \Big(  -\frac{1}{2} g_{\alpha \beta} \delta g^{\alpha \beta} (g^{\mu \lambda} \overline{\Gamma}{}^{\kappa}_{\mu \lambda} - g^{\kappa \lambda} \overline{\Gamma}{}^{\mu}{}_{\mu \lambda}) + \delta g^{\mu \lambda} \overline{\Gamma}{}^{\kappa}_{\mu \lambda} - \delta g^{\kappa \lambda} \overline{\Gamma}{}^{\mu}_{\mu \lambda} \Big) \Bigg] \\
+ \frac{1}{\sqrt{-g}} \partial_{\kappa} \Big( \sqrt{-g} g^{\mu \lambda} \delta \overline{\Gamma}{}^{\kappa}_{\mu \lambda} - \sqrt{-g} g^{\kappa \lambda} \delta \overline{\Gamma}{}^{\mu}_{\mu \lambda} \Big) \,,
\end{multline}
where all but the first term require integration by parts. It is clear that if $f$ is a linear function of $\overline{\ourB}$ then this term is a pure boundary term and does not contribute to the equations of motion. Plugging these into the variations~(\ref{Appendix_var}) and discarding boundary terms we obtain 
\begin{multline}
  \int \sqrt{-g} f_{, \overline{\ourB}} \delta  \overline{\ourB}  \, d^4x =
  \int \sqrt{-g} \Bigg\{ \delta g^{\alpha \beta} \Big[\frac{1}{2} g_{\alpha \beta}
    f_{, \overline{\ourB}} \overline{\ourB} + (\partial_{\kappa} f_{, \overline{\ourB}})
    \Big(\frac{1}{2} g_{\alpha \beta} (g^{\mu \lambda} \overline{\Gamma}{}^{\kappa}_{\mu \lambda} -
    g^{\kappa \lambda} \overline{\Gamma}{}^{\mu}_{\mu \lambda}) -
    \overline{\Gamma}{}^{\kappa}_{\alpha \beta} + \delta^{\kappa}_{\alpha}
    \overline{\Gamma}{}^{\mu}_{\mu \beta} \Big) \Big]  \\ +
  \delta \overline{\Gamma}{}^{\kappa}_{\mu \nu} \partial_{\lambda}
  f_{,\overline{\ourB}}( \delta^{\mu}_{\kappa} g^{\lambda \nu} - \delta^{\lambda}_{\kappa} g^{\mu \nu}) \Bigg\} d^4x \,. 
\end{multline}
The final terms on the first line are exactly the terms found in~(\ref{E_verify}), with the opposite signs, so we can write this as
\begin{align}
  \label{A_B_final}
  \int \sqrt{-g} f_{, \overline{\ourB}} \delta  \overline{\ourB}\, d^4x =
  \int \sqrt{-g} \Big[\frac{1}{2}\delta g^{\alpha \beta}
    \Big(g_{\alpha \beta} f_{, \overline{\ourB}} \overline{\ourB} -
    \partial_{\kappa} f_{, \overline{\ourB}}  \overline{E}_{\alpha \beta}{}^{\kappa} \Big)
    + 2 \delta \overline{\Gamma}{}^{\kappa}_{\mu \nu} \partial_{\lambda}
    f_{, \overline{\ourB}}  g^{\nu [\lambda} \delta^{\mu]}_{\kappa} \Big] d^4x \,.
\end{align}

Combining these results~(\ref{A_G_final}) and~(\ref{A_B_final}), the modified field equations are
\begin{multline}
  \delta S = \frac{1}{2\kappa} \int \sqrt{-g} \bigg\{ \delta g^{\mu \nu}
  \Big[ -\frac{1}{2} g_{\mu \nu} f + f_{, \overline{\ourG}} (\overline{G}_{\mu \nu} +
    \frac{1}{2} g_{\mu \nu} \overline{\ourG}) +
    \frac{1}{2} \overline{E}_{\mu \nu}{}^{\lambda} (\partial_{\lambda}  f_{, \overline{\ourG}} -
    \partial_{\lambda}  f_{, \overline{\ourB}}) +
    \frac{1}{2} g_{\mu \nu}  f_{, \overline{\ourB}} \overline{\ourB} \Big] \\
  + \delta \overline{\Gamma}{}^{\kappa}_{\mu \nu} \Big( P^{\mu \nu}{}_{\kappa}  f_{,\overline{\ourG}} +
  2 \partial_{\lambda} f_{, \overline{\ourB}} g^{\nu [\lambda} \delta^{\mu]}_{\kappa}  \Big) \bigg\} d^4x \,,
\end{multline}
which are given in~(\ref{full_variation}).

\section{Projective transformations}
\label{appendix_projective}

Let us state here how the generalised projective transformation
\begin{align}
  \label{Projective}
  \overline{\Gamma}{}^{\gamma}_{\alpha \beta} \rightarrow \overline{\Gamma}{}^{\gamma}_{\alpha \beta} + c_1 \delta_{\alpha}^{\gamma} P_{\beta} + c_2 \delta_{\beta}^{\gamma} P_{\alpha} \,,
\end{align}
acts on the relevant affine quantities. The Riemann tensor transforms as
\begin{align}
\overline{R}_{\nu \mu \lambda}{}^{\gamma} &\rightarrow \overline{R}_{\nu \mu \lambda}{}^{\gamma} + 2 c_2 \delta^{\gamma}_{\lambda} \partial_{[\nu} P_{\mu]} + 2 c_1  \delta_{[\mu}^{\gamma}  \partial_{\nu]}P_{\lambda} + 2 c_1 \overline{\Gamma}{}^{\gamma}_{[\nu \mu]} P_{\lambda} + 2 c_1 \overline{\Gamma}{}^{\rho}_{[\mu| \lambda} \delta_{\nu]}^{\gamma} P_{\rho} + 2 c_1^2 \delta_{[\nu}^{\gamma}P_{\mu]} P_{\lambda} \nonumber \\
&= \overline{R}_{\nu \mu \lambda}{}^{\gamma} +  2 c_2 \delta^{\gamma}_{\lambda} \partial_{[\nu} P_{\mu]} + 2 c_1 \overline{\Gamma}{}^{\gamma}_{[\nu \mu]} P_{\lambda} + 2 c_1 \delta^{\gamma}_{[\mu} \nabla_{\nu]} P_{\lambda} + 2 c_1^2 \delta_{[\nu}^{\gamma}P_{\mu]} P_{\lambda} \,,
\end{align}
 For the Levi-Civita connection this becomes
\begin{align}
R_{\nu \mu \lambda}{}^{\gamma} &\rightarrow R_{\nu \mu \lambda}{}^{\gamma} +  2 c_2 \delta^{\gamma}_{\lambda} \partial_{[\nu} P_{\mu]}  + 2 c_1 \delta^{\gamma}_{[\mu} \stackrel{\{\} }{\nabla}_{\nu]} P_{\lambda} + 2 c_1^2 \delta_{[\nu}^{\gamma}P_{\mu]} P_{\lambda} \nonumber \\
&=  R_{\nu \mu \lambda}{}^{\gamma}+  2 c_2 \delta^{\gamma}_{\lambda} \stackrel{\{\}}{\nabla}_{[\nu} P_{\mu]}  + 2 c_1 \delta^{\gamma}_{[\mu} \stackrel{\{\} }{\nabla}_{\nu]} P_{\lambda} + 2 c_1^2 \delta_{[\nu}^{\gamma}P_{\mu]} P_{\lambda} \,.
\end{align}
Defining the tensor $P_{\mu \nu} := P_{\mu} P_{\nu} - \nabla_{\mu} P_{\nu}$ and setting $c_1 = c_2 = 1$, the above (Levi-Civita) transformation can be expressed as
\begin{align}
R_{\nu \mu \lambda}{}^{\gamma} \rightarrow R_{\nu \mu \lambda}{}^{\gamma} -2 \delta^{\gamma}_{\lambda} P_{[\nu \mu]} + 2 \delta^{\gamma}_{[\nu} P_{\mu] \lambda} \,,
 \end{align}
which matches that found commonly in the literature~\cite{JS1954}.

The Ricci tensor and Ricci scalar for a general connection transform as
\begin{align}
\overline{R}_{\mu \lambda} &\rightarrow \overline{R}_{\gamma \mu \lambda}{}^{\gamma} +  2 c_2 \delta^{\gamma}_{\lambda} \partial_{[\gamma} P_{\mu]} + 2 c_1 \overline{\Gamma}{}^{\gamma}_{[\gamma \mu]} P_{\lambda} + 2 c_1 \delta^{\gamma}_{[\mu} \nabla_{\gamma]} P_{\lambda} + 2 c_1^2 \delta_{[\gamma}^{\gamma}P_{\mu]} P_{\lambda} \nonumber \\
&= \overline{R}_{\mu \lambda} +  2 c_2 \partial_{[\lambda} P_{\mu]} + 2 c_1 \, \overline{\Gamma}{}^{\gamma}_{[\gamma \mu]} P_{\lambda}   + c_1 (1-n) \nabla_{\mu} P_{\lambda} + c_1^2 (n-1) P_{\mu} P_{\lambda} \ , \\
\overline{R} &\rightarrow g^{\mu \lambda} \overline{R} _{\mu \lambda} + g^{\mu \lambda} (2 c_2 \, \partial_{[\lambda} P_{\mu]} + 2 c_1 \overline{\Gamma}{}^{\gamma}_{[\gamma \mu]} P_{\lambda}   + c_1 (1-n) \nabla_{\mu} P_{\lambda} + c_1^2 (n-1) P_{\mu} P_{\lambda}) \nonumber \\
&=\overline{R}  + 2 c_1 \overline{\Gamma}{}^{\gamma}_{[\gamma \mu]} P^{\mu} +  c_1 (1-n) \nabla^{\mu} P_{\mu} + c_1^2 (n-1) P^{\mu} P_{\mu} \ ,
\end{align}
and for the Levi-Civita connection this reduces to
\begin{align}
R_{\mu \lambda} &\rightarrow  R_{\mu \lambda}  +  2 c_2  \stackrel{\{\}}{\nabla}_{[\lambda} P_{\mu]} + c_1 (1-n) \stackrel{\{\}}{\nabla}_{\mu} P_{\lambda} + c_1^2 (n-1) P_{\mu} P_{\lambda}  \,,\\
\label{Projective_R}
R &\rightarrow R +  c_1 (1-n) \stackrel{\{\}}{\nabla}{}^{\mu} P_{\mu} + c_1^2 (n-1) P^{\mu} P_{\mu} \,.
\end{align}
Importantly, we see that for projective transformations of the form~(\ref{Projective}) with $c_1=0$ the Ricci scalar is invariant.

Next we have the Palatini decomposition of the affine Ricci scalar $\overline{R}= \overline{\mathbf{G}} + \overline{\mathbf{B}}$, where $\overline{\ourG}$ and $\overline{\ourB}$ are defined in~(\ref{G}) and~(\ref{B}), which we also state here for clarity
\begin{align}
  \overline{\mathbf{G}} &= g^{\mu \lambda} \big(  \overline{\Gamma}{}^{\kappa}_{\kappa \rho}  \overline{\Gamma}{}^{\rho}_{\mu \lambda} -  \overline{\Gamma}{}^{\kappa}_{\mu \rho}  \overline{\Gamma}{}^{\rho}_{\kappa \lambda} \big) 
  + \big( \overline{\Gamma}{}^{\mu}_{\mu \lambda} \delta^{\nu}_{\kappa} -  \overline{\Gamma}{}^{\nu}_{\kappa \lambda} \big) \big( \partial_{\nu} g^{\kappa \lambda} - \frac{1}{2} g_{\alpha \beta} g^{\kappa \lambda} \partial_{\nu} g^{\alpha \beta} \big)\,,
  \nonumber \\
  \overline{\mathbf{B}} &= \frac{1}{\sqrt{-g}} \partial_{\kappa} \big(\sqrt{-g}(g^{\mu \lambda}  \overline{\Gamma}{}^{\kappa}_{\mu \lambda} - g^{\kappa \lambda}  \overline{\Gamma}{}^{\mu}_{\mu \lambda})\big)\,.
\nonumber
\end{align}
The projective transformation for the bulk quantity $\overline{\mathbf{G}}$ is 
\begin{align}
\overline{\mathbf{G}} \rightarrow \overline{\mathbf{G}} + c_1 \Big[ 2 \overline{\Gamma}{}^{\kappa}_{[\kappa \mu]} P^{\mu} + (n-1) \big(g^{\mu \lambda}  \overline{\Gamma}{}^{\rho}_{\mu \lambda} P_{\rho} +P_{\lambda} \partial_{\nu} g^{\nu \lambda} - \frac{1}{2} P^{\nu} g_{\alpha \beta} \partial_{\nu} g^{\alpha \beta} \big)  + c_1 (n-1) P^{\lambda} P_{\lambda}
\Big] \,,
\end{align}
This can also be rewritten explicitly in terms of non-metricity and torsion\footnote{Here if one works with the Levi-Civita connection, we see only the $P^{\lambda}P_{\lambda}$ term remains. This, along with the total divergence term $\nabla_{\kappa} P^{\kappa}$  in the $\mathbf{B}$ transformation below, immediately gives the Schouten result~(\ref{Projective_R}).}
\begin{align}
  \overline{\mathbf{G}} \rightarrow \overline{\mathbf{G}} + c_1 \Big[ (2-n) T^{\kappa}{}_{\kappa \mu} P^{\mu} + (n-1) P_{\lambda}\big(\nabla_{\nu}g^{\nu \lambda} - \frac{1}{2} g_{\alpha \beta} \nabla^{\lambda} g^{\alpha \beta}\big) + c_1 (n-1) P^{\lambda} P_{\lambda} \Big] \,.
\end{align}
For the boundary term $\overline{\mathbf{B}}$ we find
\begin{align}
  \overline{\mathbf{B}} \rightarrow \overline{\mathbf{B}} + c_1 (1-n)\frac{1}{\sqrt{-g}} \partial_{\kappa}(\sqrt{-g} P^{\kappa}) \,.
\end{align}
Both of these affine quantities share the projective properties of the Ricci scalar, which are invariant for transformations with $c_1=0$. From this we can deduce that any action comprised of these objects will share this invariance, see~\cite{Sotiriou:2006qn}. Importantly, this implies that the connection variations will lead to equations that are trace free over these indices. The variation of the $f(\overline{\ourG},\overline{\ourB})$ action with respect to the connection leads to the equation
\begin{align}
  P^{\mu \nu}{}_{\kappa}  f_{, \overline{\ourG}} +
  2\partial_{\lambda} f_{,\overline{\ourB}} g^{\nu [\lambda} \delta^{\mu]}_{\kappa} \,,
\end{align}
with $P^{\mu \nu}{}_{\kappa}$ the Palatini tensor~(\ref{Palatini}). This is easily verified to be trace free over the indices $\nu \, \kappa$
\begin{multline}
  P^{\mu \nu}{}_{\nu} f_{,\overline{\ourG}} +2\partial_{\lambda} f_{, \overline{\ourB}} g^{\nu [\lambda} \delta^{\mu]}_{\nu} =
  f_{,\overline{\ourG}} \Bigl(-Q_{\nu}{}^{\mu \nu} + \frac{1}{2} g^{\mu \nu} Q_{\nu \lambda}{}^{\lambda} +
  \delta_{\nu}^{\mu} Q_{\lambda}{}^{\lambda \nu} -
  \frac{1}{2} \delta^{\mu}_{\nu} Q^{\nu}{}_{\lambda}{}^{\lambda} \\
  + T^{\mu}{}_{\nu}{}^{\nu} + g^{\mu \nu} T^{\lambda}{}_{\lambda \nu} +
  \delta^{\mu}_{\nu} T^{\lambda \nu}{}_{\lambda}\Bigr) = 0 \,,
\end{multline}
where all non-metricity terms and torsion terms cancel due to their symmetry properties. 

For consistency, we show that the above transformations give the correct transformation for the Ricci scalar. First, expanding the boundary term transformation gives
\begin{align}
\overline{\mathbf{B}} \rightarrow \overline{\mathbf{B}} + c_1 (1-n)\big(\partial_{\kappa} P^{\kappa}-\frac{1}{2} g_{\alpha \beta} P^{\kappa} \partial_{\kappa} g^{\alpha \beta} ) \,,
\end{align}
which then leads to
\begin{align}
\overline{\mathbf{G}} + \overline{\mathbf{B}} &\rightarrow \overline{\mathbf{G}} + \overline{\mathbf{B}}   + c_1 \Big[ 2\overline{\Gamma}{}^{\kappa}_{[\kappa \mu]} P^{\mu} + c_1 (n-1) P^{\lambda} P_{\lambda} \nonumber
 \\
&+ (n-1) \big(g^{\mu \lambda} \overline{\Gamma}{}^{\rho}_{\mu \lambda} P_{\rho} + P_{\lambda} \partial_{\nu} g^{\nu \lambda} - \frac{1}{2} P^{\nu} g_{\alpha \beta} \partial_{\nu} g^{\alpha \beta} + \frac{1}{2} g_{\alpha \beta} P^{\kappa} \partial_{\kappa} g^{\alpha \beta} - \partial_{\kappa} P^{\kappa} ) 
\Big]  \nonumber \\
&= \overline{\mathbf{G}} +\overline{\mathbf{B}}  + c_1 \Big[ 2\Gamma^{\kappa}_{[\kappa \mu]} P^{\mu} + c_1 (n-1) P^{\lambda} P_{\lambda} + (n-1) \big(g^{\mu \lambda} \overline{\Gamma}{}^{\rho}_{\mu \lambda} P_{\rho} - \partial^{\kappa} P_{\kappa} ) 
\Big] \nonumber \\
&=  \overline{\mathbf{G}} +\overline{\mathbf{B}}  + 2 c_1 \overline{\Gamma}{}^{\kappa}_{[\kappa \mu]} P^{\mu} +  c_1^2 (n-1) P^{\lambda} P_{\lambda} + c_1 (1-n)  \nabla^{\mu} P_{\mu} \,,
\end{align}
which matches the previous calculations.

\section{Cosmological solution}
\label{appODE}

The ODE~(\ref{ODE}) was given by
\begin{align}
  \frac{\dot{a}}{a} = \frac{\sqrt{2}}{2\sqrt{3\beta}} \frac{Y\sqrt{Y^2+2Y-3}}{Y^2+4Y-1}\,,
  \qquad Y^2 = 1+\frac{36\beta\rho_0}{a^3} \,.
  \label{a1ODE}
\end{align}
We begin by introducing a new unknown function as follows
\begin{align}
  \sinh(y(t)) = \sqrt{\frac{36\beta\rho_0}{a^3}}\,,
  \label{a2ODE}
\end{align}
which simplifies the ODE substantially due to various hyperbolic identities. One finds
\begin{align}
  \frac{dy}{dt} = -\frac{1}{\sqrt{3\beta}}\frac{\sqrt{3+\cosh(y)}\sech(y/2)}{2+8\coth(y)\csch(y)} \,.
  \label{a3ODE}
\end{align}
After separation of variables and integration this gives
\begin{align}
  \sqrt{3\beta/2}(t-t_0) = \sqrt{\frac{\cosh(y)+3}{\cosh(y)-1}} -
  2\arctan\Bigl(\frac{\sqrt{2}\sinh(y/2)}{\sqrt{3+\cosh(y)}}\Bigr) -
  2\arctanh\Bigl(\frac{1}{\sqrt{2}}\sinh(y/2)\Bigr) \,.
\end{align}
Here $t_0$ is the constant of integration. Using the expression of $y$ in terms of $a$ gives an implicit formula for $a(t)$.

\addcontentsline{toc}{section}{References}

\providecommand{\href}[2]{#2}\begingroup\raggedright\endgroup

\end{document}